# Brain-aligning of semantic vectors improves neural decoding of visual stimuli

Shirin Vafaei[1], Ryohei Fukuma[1,2], *Takufumi Yanagisawa[1,2,3], Huixiang Yang[2], Satoru Oshino[1,3], Naoki Tani[1,3], Hui Ming Khoo[1,3], Hidenori Sugano[4], Yasushi Iimura[4], Hiroharu Suzuki[4], Madoka Nakajima[4], Kentaro Tamura[5,6], Haruhiko Kishima[1]

[1]Department of Neurosurgery, Graduate School of Medicine, University of Osaka, Suita, Japan

[2]Department of Neuroinformatics, University of Osaka Graduate School of Medicine, Suita, Japan

[3]Osaka University Hospital Epilepsy Center, Suita, Japan

[4]Department of Neurosurgery, Juntendo University, Tokyo, Japan

[5]Department of Neurosurgery, Nara Medical University, Kashihara, Japan

[6]Department of Neurosurgery, National Hospital Organization, Nara Medical Center, Nara, Japan

***Corresponding author**

Takufumi Yanagisawa

Address: 2-2 Yamadaoka, Suita, Osaka 565-0871, Japan

Tel.: +81-6-6879-3652

Fax: +81-6-6879-3659

E-mail: tyanagisawa@nsurg.med.osaka-u.ac.jp

## Abstract

The development of algorithms to accurately decode neural information has long been a research focus in the field of neuroscience. Brain decoding typically involves training machine learning models to map neural data onto a preestablished vector representation of stimulus features. These vectors are usually derived from image- and/or text-based feature spaces. Nonetheless, the intrinsic characteristics of these vectors might fundamentally differ from those that are encoded by the brain, limiting the ability of decoders to accurately learn this mapping. To address this issue, we propose a framework, called brain-aligning of semantic vectors, that fine-tunes pretrained feature vectors to better align with the structure of neural representations of visual stimuli in the brain. We trained this model with functional magnetic resonance imaging (fMRI) and then performed zero-shot brain decoding on fMRI, magnetoencephalography (MEG), and electrocorticography (ECoG) data. fMRI-based brain-aligned vectors improved decoding performance across all three neuroimaging datasets when accuracy was determined by calculating the correlation coefficients between true and predicted vectors. Additionally, when decoding accuracy was determined via stimulus identification, this accuracy increased in specific category types; improvements varied depending on the original vector space that was used for brain-alignment, and consistent improvements were observed across all neuroimaging modalities.

## Introduction

The development of brain decoding algorithms is essential for advancing brain–machine interfaces (BMIs)[1–4] that enable precise communication and motor control for individuals with speech or motor impairments. Moreover, these algorithms offer a unique opportunity to explore the complexities and fundamental mechanisms underlying information processing in the human brain[5–10]. Furthermore, accurate decoding can improve the effectiveness of neurofeedback systems by enabling the decoding of cognitive patterns and delivering real-time neurofeedback, thereby assisting patients in the refinement of their cognitive and emotional faculties[11–14].

Previous studies have shown that neural activity patterns can be decoded to reveal information about perceived or imagined visual stimuli (i.e., images). This information can take the form of semantic attributes[10,15–18], category-level classes[5,7,15,19,20] or even reconstructed visual representations of the images[21–24]. Decoding typically involves representing a specified attribute as a pretrained feature vector, often derived from object recognition neural networks[15], multimodal models[25,26] or word co-occurrence statistics[12,27–29]. Then, machine learning models are trained to map neural activity patterns to these feature vectors.

While pretrained feature vectors have enabled brain decoding with above-chance accuracy, current models are still limited in accurately learning this mapping, particularly in zero-shot decoding scenarios, where decoders must generalize the learned information to novel semantic categories not encountered during training[30]. Given the impracticality of training decoders to learn the representations of all possible semantic categories, developing more robust and flexible decoding models has become imperative.

In this study, we hypothesized that if the vectors that are used to represent stimuli are more aligned with how visual stimuli are encoded in the human brain, decoders can better learn the mapping of neural activity patterns to feature vectors and even generalize this mapping to novel semantic categories using learned information encapsulated within the more brain-aligned vectors.

This idea was inspired by recent findings that the use of brain-like or brain-integrated features can improve object recognition[31], few-shot learning and anomaly detection tasks[32] and that consistent and high-performing latent spaces can be obtained by jointly learning from both behavioral and neural data[33]. However, whether semantic spaces with representations that are more aligned with neural encoding patterns can lead to more accurate zero-shot brain decoding remains unclear.

To create a brain-aligned semantic vector representation of stimuli, we propose a framework called *brain-aligning of semantic vectors*, which reconstructs the pretrained feature vectors while ensuring that the second-order statistical features of its latent space are as similar as possible to those of brain activity patterns. The vectors that are extracted from the latent space of the autoencoder after training are called *brain-aligned semantic vectors*. We investigated whether utilizing these brain-aligned semantic vectors can improve zero-shot decoding accuracy to identify the predicted vector from a set of candidate vectors[15,34] for (1) brain activity measured by the same neuroimaging technique used to fine-tune the pretrained feature spaces and (2) brain activity measured by a different neuroimaging technique than that used for fine-tuning.

Specifically, we trained the brain-aligning framework by leveraging brain activity patterns measured by functional magnetic resonance imaging (fMRI) and then tested the zero-shot decoding performance of the resulting brain-aligned vectors on brain activity data measured by fMRI, magnetoencephalography (MEG) and electrocorticography (ECoG). This cross-modality approach is critical because fMRI, MEG and ECoG measure distinct aspects of brain activity: fMRI captures hemodynamic changes (BOLD signals)[35], ECoG records electrical activity[36], and MEG detects magnetic fields[37]. Successful generalization across modalities would suggest that our vectors represent fundamental aspects of neural coding that are independent of specific measurement techniques.

## Results

### Brain-aligning of semantic vectors

We developed a multimodal learning autoencoder framework that takes pretrained feature vectors and the brain activity patterns corresponding to the visual stimuli dataset and aligns the feature space with the structure of visual representations in the human brain. To select the pretrained feature vectors, we used two different feature spaces. The first is an image-based feature space that includes features from the image encoder model of CLIP[26], and the second is a text-based feature space that includes features from the global vectors for word representation (GloVe) model[29]. The fMRI dataset that was used here to fine-tune the pretrained feature vectors was the generic object decoding (GOD) dataset[15]. This dataset contains fMRI recordings from 5 subjects viewing 1200 images of 200 distinct object categories that were selected from ImageNet[38]. The GOD dataset has been specifically designed to prevent any overlap between the categories used for training and those used for testing, facilitating assessment of the zero-shot prediction capabilities of decoding models. Importantly, the use of this dataset ensures that these models are evaluated on the basis of their ability to generalize to entirely new categories without prior exposure.

First, we extracted the original feature vectors for each category in the GOD dataset (see Methods) and represented each category by its corresponding feature vector. To obtain the brain-aligned semantic space, we trained the autoencoder with a two-term objective function. The first term is a simple mean squared error (MSE) loss between true and predicted pretrained feature vectors (with the goal of reconstructing them). The second term is the MSE loss between the representational similarity matrix (RSM)[39] of the fMRI signals and the autoencoder's latent space in each batch. Mathematically:

$$loss = \frac{1}{m}\sum_{i=1}^{m}\left[(\alpha)\left(y - y^{\prime}\right)^{2} + (1-\alpha)(RSM_l - RSM_b)^2\right] \quad (1)$$

where $y$ is the original semantic vector, $y'$ is the reconstructed semantic vector, $RSM_l$ is the RSM of the autoencoder's hidden layer, $RSM_b$ is the RSM of the corresponding brain activity patterns, and $m$ is the number of samples in each batch. Finally, $\alpha$ is the hyperparameter that determines the extent of brain alignment.

Given our aim to decode visual object categories, we used the fMRI data from different brain regions of interest (ROIs) in the visual cortex (lower visual areas (V1–V4), the lateral occipital complex (LOC)[40], the fusiform face area (FFA)[41], and the parahippocampal place area (PPA)[42]; see the Methods section for the definition of the ROIs). Concurrently, to choose the appropriate pretrained feature vectors, we extracted the category-specific pretrained feature vectors of each category in the training data of the GOD dataset.

For each combination of ROIs, participants in the GOD dataset, pretrained feature vectors and α values, we trained a different autoencoder using a leave-one-subject-out procedure. Specifically, when creating brain-aligned vectors for decoding a particular subject (e.g., Subject A), the autoencoder was trained using (1) the pretrained semantic vectors of all training categories from the GOD dataset, and (2) the averaged RSMs computed from the other 4 subjects' fMRI brain signals for these same training categories. This leave-one-subject-out approach ensures that the brain-aligned semantic space is not influenced by Subject A's idiosyncratic neural response patterns, thereby preventing information leakage and ensuring that the semantic space captures generalizable neural structure across individuals. We trained autoencoders on a wide range of values of $\alpha$ ($\alpha$ = 0.0001, 0.001, 0.01, 0.1, and 1). We specifically included 1 as one of the values to determine how excluding the brain-aligning part from this framework would affect the downstream analyses. Figure 1 shows an overview of the proposed workflow.

**fMRI brain decoding of visual stimuli**

We performed brain decoding for each ROI, feature space type and participant separately. For each subject (e.g., Subject A), linear regression decoders were trained to map that subject's brain activity patterns from the training categories to their corresponding brain-aligned semantic vectors (which were created without using Subject A's data, as described above). The trained decoders were then evaluated on Subject A's brain activity for the test categories, enabling zero-shot generalization to novel semantic categories not encountered during decoder training. Importantly, for each ROI (e.g., V4), both the autoencoder and decoder were trained and tested exclusively on the voxel activity from that same ROI, using the brain-aligned vectors derived from that ROI’s autoencoder. Across all samples and for each unit in the semantic vectors, a separate set of linear regression models was trained. We evaluated the identification accuracies of the models to assess their ability to correctly classify stimuli on the basis of the predicted feature vectors. This evaluation is critical for determining the practical utility of decoding models used in real-world applications, such as brain-machine interfaces or neurofeedback systems[6,34]. For this purpose, we computed the Pearson correlation coefficient between the predicted semantic vector and all the other candidate vectors. The accuracy is defined as the percentage of candidate categories whose correlation with the predicted vector is lower than the correlation between the true and predicted vectors. The final identification accuracy is determined by averaging the identification accuracies across all categories and subjects in the test dataset. Figure 2 shows an overview of the brain identification algorithm.

We first evaluated the performance of category identification relative to chance. Specifically, we compared the identification results obtained from the original data with those obtained from shuffled data. The shuffled accuracy was obtained by correlating the predicted vectors with both the shuffled true vector and all the other candidate vectors. The shuffled identification accuracy was then defined as the percentage of candidate categories whose correlation with the predicted vector was lower than the correlation between the shuffled true and predicted vectors. The real identification accuracies were significantly greater than the corresponding shuffled accuracies for 42/42 of the CLIP perception data, 35/42 of the CLIP imagery, 42/42 of the GloVe perception and 42/42 of the GloVe imagery (one-sided *t*-test, $p < 0.05$, see Supplementary Tables 1–4 for the exact values).

The identification results revealed that brain-aligned semantic vectors can enhance zero-shot visual stimulus identification across the visual cortical hierarchy. The higher-order visual areas, particularly V4, LOC, FFA, and PPA, consistently outperformed the early visual areas (V1, V2, V3) in terms of the identification accuracy, reflecting their specialized role in object recognition and semantic processing (Fig. 3; see Supplementary Figure 1 and Supplementary Tables 5–8 for the comprehensive statistical results and individual subject data). Importantly, the optimal degree of brain alignment varied depending on the semantic feature space used. For the CLIP-based vectors, moderate brain alignment ($\alpha = 0.1$) yielded the best identification performance, whereas the GloVe-based vectors benefited most from slightly stronger alignment ($\alpha = 0.01$). Both feature types showed peak performance when aligned with V4 neural activity patterns, which is consistent with the demonstrated specialization of V4 for object-like shape processing[43,44]. The identification accuracies remained robust across both the perception and the imagery conditions, although the performance of the imagery condition was slightly lower overall. These findings demonstrate that there exists an optimal balance for brain alignment—neither too weak nor too strong—that maximizes the model's ability to correctly identify visual stimuli in zero-shot scenarios and that this optimal alignment is best captured by intermediate visual areas that balance perceptual detail with semantic abstraction.

**Generalization of the decoding performance to other modalities: MEG and ECoG datasets**

To assess the generalizability of using the fMRI-derived brain-aligned vectors to decode other types of neuroimaging brain data, we performed decoding analysis on MEG and ECoG neural data from different participants who were exposed to the same visual stimuli.

We trained separate linear regression models as brain decoders for each of the MEG and ECoG datasets, subjects and feature space types. The decoders used brain-aligned vectors that were derived from fMRI V4 data (which showed optimal performance in the fMRI experiments). For MEG, the decoders were trained on source-estimated signals from the ventral visual stream regions; for ECoG, the decoders were trained on high-$\gamma$ power from subdural electrodes covering the ventral visual cortex.

Figure 4 shows the identification results of the MEG neural data using the fMRI brain-aligned vectors obtained from the V4 brain region. We first evaluated whether the MEG neural data could be successfully decoded using the original and fMRI-derived brain-aligned vectors from the V4 region by comparing identification accuracies against shuffled data using one-sided *t*-tests. For the CLIP-based vectors, all conditions achieved significantly above-chance performance ($p < 0.001$, one-sided *t*-test, Fig 4a, see Supplementary Table 9 for the exact values). In contrast, for the GloVe-based vectors, the original vectors failed to achieve above-chance performance ($p = 0.446$), whereas brain-aligned vectors consistently exceeded chance levels ($p < 0.001$, see Supplementary Table 10 for the exact values). These results demonstrate that brain alignment transforms originally ineffective GloVe vectors into successful decoders for MEG data, highlighting the importance of aligning semantic representations with brain activity patterns for cross-modal generalization. Next, we examined differences in identification accuracy between the original, vectors with $\alpha = 1$, and the optimal brain-aligned vectors from the fMRI results. The MEG identification results demonstrate that the brain-aligned semantic vectors derived from the fMRI data effectively transfer to a different neuroimaging modality. For the CLIP-based vectors, the optimal brain-aligned condition ($\alpha = 0.1$) substantially outperformed both the original pretrained vectors and the reconstruction-only condition ($\alpha = 1$). With respect to the GloVe-based vectors, compared with the original vectors, both the reconstruction-only ($\alpha = 1$) and the brain-aligned conditions ($\alpha = 0.01$) resulted in enhanced performance, with the brain-aligned conditions achieving higher identification accuracies. Importantly, the brain-alignment parameters that proved optimal in the original fMRI training ($\alpha = 0.1$ for CLIP and $\alpha = 0.01$ for GloVe) maintained their superior performance when applied to MEG data. For the comprehensive statistical results and subject-by-subject results, see Supplementary Figures 2–3 and Supplementary Tables 11–12.

Similarly, we first evaluated whether ECoG neural data could be successfully decoded using the original and fMRI-derived brain-aligned vectors from the V4 region by comparing identification accuracies against shuffled data. For the CLIP-based vectors, brain-aligned conditions with $\alpha = 1$ and $\alpha = 0.1$ achieved significantly above-chance performance ($p = 0.0029$ and $p < 0.001$, respectively, one-sided $t$-test, Fig 5a, see Supplementary Table 13 for the exact values), whereas the original vectors failed to reach significance ($p = 0.9991$). For the GloVe-based vectors, all the conditions achieved significantly above-chance performance: original vectors ($p = 0.0007$), $\alpha = 1$ vectors ($p < 0.001$), and $\alpha = 0.01$ vectors ($p < 0.001$, Fig 5b, see Supplementary Table 14 for the exact values).

Next, we examined differences in identification accuracy between the original vectors, vectors with $\alpha = 1$, and the optimal brain-aligned vectors identified from the fMRI results. The ECoG identification results demonstrate that the brain-aligned semantic vectors derived from the fMRI data effectively transfer to invasive neural recordings. For the CLIP-based vectors, the optimal brain-aligned condition ($\alpha = 0.1$) substantially outperformed both the original pretrained vectors and the reconstruction-only condition ($\alpha = 1$). With respect to the GloVe-based vectors, the performance of the brain-aligned condition ($\alpha = 0.01$) was better than that of the original vectors. While the group-level analysis revealed comparable performance between the brain-aligned vectors with $\alpha = 0.01$ and the vectors with $\alpha = 1$, the individual subject analysis revealed that 3 out of 4 subjects achieved higher identification accuracy with $\alpha = 0.01$ than with $\alpha = 1$. Importantly, the brain-alignment parameters that proved optimal in the original fMRI training ($\alpha = 0.1$ for CLIP and $\alpha = 0.01$ for GloVe) maintained their superior performance when applied to the ECoG data. These findings highlight the robust cross-modal transferability of brain-aligned semantic representations across different neuroimaging modalities (see Supplementary Figures 4–5 and Supplementary Tables 15–16 for the subject-by-subject results).

**Consistency of the identification accuracy improvement among neuroimaging modalities**

We investigated whether brain alignment produced category-specific changes in the identification accuracy and tested whether these changes were consistent across neuroimaging modalities. We first clustered the categories in ImageNet into 10 clusters in both the original CLIP and GloVe space separately (see Supplementary Figures 6–7 for the clustering results and procedures). For each optimal $\alpha$ in each neuroimaging modality, original semantic space (all derived from fMRI V4), and category, we calculated the difference in the identification accuracy between the aligned vectors and the original vectors. Figure 6 shows the improvement and visualization of the categories in the clusters for CLIP, and Figure 7 shows the improvement and visualization of the categories in the clusters for GloVe. The brain-alignment effects varied by category type: the CLIP vectors showed greater improvements for artifacts and object categories (e.g., vehicles, tools), whereas the GloVe vectors demonstrated preferential enhancement for biological categories (e.g., animals), indicating that visual-semantic and text-based representations align differently with brain activity patterns across semantic domains (see Supplementary Figures 8–13).

**Comparison of category discriminability between the original and brain-aligned CLIP features**

To investigate whether brain alignment affects the intrinsic categorical structure of semantic representations, we analyzed the category discriminability of individual feature units, defined as the *F* statistic measuring the ratio of intercategory to intracategory variation in the feature values[15]. Since the GloVe vectors are identical for all images within a category, this analysis was restricted to the CLIP-based feature vectors. We computed discriminability metrics for each of the feature units across 19,933 ImageNet categories, with 8 images per category, and compared the original CLIP vectors against the brain-aligned variants that were trained on the fMRI V4 neural patterns with different alignment strengths. A direct paired statistical comparison between the original (512-dimensional) and brain-aligned (256-dimensional) features was not feasible because of unequal numbers of feature units for the Wilcoxon signed-rank test. Instead, we compared the brain-aligned conditions ($\alpha \leq 0.1$) against the $\alpha = 1$ baseline, which represents autoencoder-compressed features without neural constraints. All the brain-aligned conditions significantly outperformed this baseline (all $p < 0.005$, Wilcoxon signed-rank test, Fig 8), indicating that neural constraints improve feature representations beyond the effects of dimensionality reduction.

## Discussion

Here, we demonstrated the ability of our proposed brain-aligning method to enhance zero-shot brain decoding across diverse neuroimaging datasets and distinct individuals. Notably, the fMRI brain decoders that were trained on the CLIP-based and GloVe-based brain-aligned feature vectors outperformed those that were trained on the original pretrained vectors (Fig. 3). Importantly, this improvement was observed even when other types of neuroimaging neural data (MEG and ECoG), subjects, and stimulus categories that were not included in training the brain-aligning model were considered (Figs. 4 and 5), highlighting the generalizability of our approach.

Previous studies have attempted to develop a brain-based semantic representation space. For example, Binder *et al*.[45] proposed a model in which word meanings are represented as combinations of basic sensory, motor, affective, and cognitive experiences. These authors introduced a basic set of approximately 65 experiential attributes on the basis of neurobiological considerations—spanning sensory, motor, spatial, temporal, affective, social, and cognitive domains—and collected normative data on these experiential attributes to create a semantic space based on brain activity. In another study, Chersoni *et al*.[46] advanced this concept by demonstrating the decoding of word embeddings using brain-based semantic features proposed by Binder *et al*. However, while Binder *et al*. established a foundation for linking brain activity to semantic representations, their approach relied on manually-defined attributes rather than directly using the raw neural data. In contrast, our approach harnesses the inherent structure of neural representations by directly using the second-order statistical characteristics of brain activity patterns, avoiding the need to manually define attributes. This data-driven approach may offer a more direct and potentially comprehensive representation of neural semantic space.

In addition to creating brain-based semantic spaces, as exemplified by Binder *et al*., several studies have fine-tuned neural representations with human, monkey or rat brain data to enhance performance on downstream tasks. For example, Federer *et al*.[31] reported that training neural networks to mimic the statistical properties of brain activity can improve object recognition. Later, Li *et al*.[47] integrated deep neural network features with brain network information to enhance the prediction of brain activity during naturalistic perception. Additionally, Muttenthaler *et al*.[32] explored aligning neural network representations with human similarity judgments to improve few-shot learning and anomaly detection. Finally, Schneider *et al*.[33] demonstrated the power of combining behavioral and neural data through latent embeddings for predicting behavior. However, despite these advancements, these previous studies did not explicitly explore the fine-tuning of pretrained feature vectors to directly match the second-order statistical representations of human brain activity, nor did they systematically investigate the resulting zero-shot decoding performance on new subjects and neuroimaging modalities. Our brain-aligning method addresses this gap by aligning feature vector relationships with those observed in neural responses, demonstrating robust cross-modality and cross-subject decoding capabilities.

Our findings are also aligned with the broader literature addressing hyperalignment[48] and the need to discover shared neural representational spaces across individuals. While our primary goal was not to derive a common high-dimensional space *per se*, our results nevertheless suggest some degree of alignment across individuals. By creating brain-aligned vectors on the basis of averaged representational similarity matrices (RSMs) across subjects, we effectively leveraged the neural representations that are common across individuals. The subsequent successful decoding of neural activity patterns from a different set of subjects aligns with previous findings such as those of Guntupalli *et al*.[49] who demonstrated the feasibility of finding such shared spaces even at a fine-grained, searchlight level. Furthermore, our results build upon the notion of shared representations across neuroimaging modalities[39,50,51], which is consistent with findings from studies such as Haxby *et al*.[48] that suggest the existence of common representational structures in fMRI data. Notably, in our study, the successful decoding of MEG signals and even ECoG signals via our fMRI-derived brain-aligned vectors provides evidence for a shared representational space with consistent second-order statistical characteristics across these distinct modalities.

In conclusion, our study demonstrates the notable potential of brain-aligning semantic vectors in increasing the accuracy and generalizability of neural decoding algorithms. By integrating brain-related information into pretrained feature vectors, we improved zero-shot decoding performance across different individuals and neuroimaging modalities, even with a relatively small fMRI dataset (consisting of approximately 150 categories). This suggests that our approach efficiently captures essential neural representations even with limited training data. While these results are promising, future exploration of several issues is needed. For example, investigating the impact of different autoencoder architectures and loss metrics as well as leveraging larger datasets could further optimize the effectiveness of brain-aligning vectors. Additionally, developing methods to mitigate potential biases in the brain-aligning process and enhance the interpretability of the resulting vectors would facilitate real-world applications. In addition to these immediate refinements, future work could explore the application of brain-aligning to a broader range of cognitive domains and tasks, ultimately paving the way for more powerful and versatile brain-machine interface technologies.

## Methods

### Creating semantic vectors

Semantic vectors are multidimensional representations of data that encode the underlying semantics, relationships, and context within that data. These vectors have been widely used to decode meaningful representations of stimuli in the brain; thus, decoders are trained to map neural activity patterns to corresponding semantic vector representations. Here, we used two different types of semantic spaces that have been previously used in brain decoding studies. Specifically, we use pretrained feature vectors from the last layer of the CLIP image encoder and pretrained feature vectors from the GloVe model. We created semantic vectors for all categories in the ImageNet dataset (fall 2011 release)[38].

*GloVe*

GloVe is a method that generates 300-dimensional semantic vector representations of words based on normalized word co-occurrence statistics obtained from a corpus containing more than 42 billion tokens. Words with similar meanings are associated with vectors that are close in the vector space, enabling GloVe to capture the semantic meaning of words and their contextual associations. Here, we used the pretrained word vectors of the 42B token file (https://nlp.stanford.edu/data/glove.42B.300d.zip). For each image category in the ImageNet dataset, we used their crowdsourced annotations[38] and calculated the average GloVe representations of all available annotations in the GloVe dictionary as a representation of that category. If any of the annotations of a particular category did not exist in the GloVe dictionary, that category was excluded from all subsequent analyses.

*CLIP*

CLIP is a model that connects vision and language by encoding semantic vectors for both images and text. The unique advantage of using CLIP lies in its ability to map images and textual descriptions into a shared vector space, where the similarity or dissimilarity between vectors accurately reflects the semantic relationships between the two modalities. To create a CLIP semantic vector for each category in ImageNet, we extracted an image from each category and then extracted the features from the ViT-B/32 transformer image encoder of the CLIP model for that image.

## fMRI dataset

### *Dataset description*

We used the publicly available "Generic Object Decoding" dataset[15]. Five healthy subjects (one female and four males, aged between 23 and 38 years) with normal or corrected-to-normal vision participated in the experiments. The sample size was chosen to match previous fMRI studies with comparable research objectives. The experiments consisted of presenting natural object images to the subjects and recording their brain activity while they viewed the visual stimuli (perception experiment) or imagined them (imagery experiment). Images were selected from the ImageNet dataset (2011, fall release). The training dataset consisted of neural recordings of 1200 images (150 categories, 8 images per category), all of which were viewed by the participants. The test dataset consisted of neural recordings of 50 seen and 50 imagined images (50 images were selected from 50 categories, i.e., 1 image per category, and were not used in the training dataset; the training and test images were presented 35 and 10 times, respectively).

All the subjects provided written informed consent, and the study protocol was approved by the Ethics Committee of the ATR.

### *ROI identification and selection*

In the GOD dataset, the borders of visual cortical areas were delineated using both retinotopic mapping and functional localizer experiments. In the retinotopic mapping experiment, subjects were presented with two types of stimuli: rotating wedges and expanding rings composed of flickering checkerboards. The retinotopy data were then transformed into Talairach space, and the boundaries of visual areas V1–V4 were identified on flattened cortical surfaces using BrainVoyager QX software (http://brainvoyager.com)[19,52]. Higher-level visual regions (lateral occipital complex, LOC; fusiform face area, FFA; and parahippocampal place area, PPA) were identified through functional localizer experiments in which subjects viewed both intact and scrambled images of faces, objects, houses, and scenes[40–42].

### MEG dataset

*Subjects*

Three healthy subjects (male, aged between 25 and 34 years) with normal or corrected-to-normal vision participated in the experiments. All the participants were informed about the experiment's purpose and procedure and provided written informed consent. The study adhered to the Declaration of Helsinki and was performed in accordance with protocols approved by the Ethics Committee of Osaka University Clinical Trial Center (Protocol No. 18472-5).

*Visual images*

Visual stimuli were drawn from the GOD dataset, which comprises images collected from ImageNet (2011, fall release). The dataset contains images from 200 distinct object categories.

Images underwent square cropping preprocessing according to methods described in a previous study[15].

*MRI acquisition*

T1-weighted MRI data from each subject were collected using a 3.0-Tesla SYNAPSE VINCENT scanner (Fujifilm, Tokyo, Japan) located at Osaka University's hospital.

*MEG acquisition*

Prior to starting the MEG recordings, five marker coils were attached to the subject's face to determine the position and orientation of the MEG sensors relative to the head, and the head position was evaluated using these coils before and after each recording (maximum acceptable displacement: 5 mm). To coregister the MEG data of each subject with their corresponding MRI data, 100 points were digitized on the scalp of each participant (FastSCAN Cobra; Polhemus, Colchester, VT, USA).

The MEG signals were recorded using a 160-channel whole-head MEG system equipped with coaxial-type gradiometers housed in a magnetically shielded room (MEGvision NEO; Yokogawa Electric Corporation, Kanazawa, Japan). The subjects were placed in a supine position with their head centered on the gantry. To minimize shoulder movement artifacts, a cushion was positioned under the subject's elbows. The subjects were explicitly instructed to keep their head stationary to prevent motion artifacts.

Visual stimuli were presented using a projection screen that was located in front of the subject's face (Presentation; Neurobehavioral Systems, Albany, CA, USA) and a liquid crystal projector (LVP-HC6800; Mitsubishi Electric, Tokyo, Japan). The MEG signals were passed through an optical isolation circuit and sampled at 1000 Hz with an online 200 Hz low-pass filter using FPGA DAQ boards (PXI-7854R; National Instruments, Austin, TX, USA).

*Experimental design*

For all the images in the GOD visual stimuli dataset, we conducted an image presentation experiment. All the visual stimuli were rear-projected onto a screen in the MEG scanner bore using a luminance-calibrated liquid crystal display projector (LVP-HC6800; Mitsubishi Electric, Tokyo, Japan). Data from each subject were collected over multiple scanning sessions. On each experimental day, one session was conducted for a maximum of 1 hour. Each session included two types of runs: “rest” runs (first and last runs, not counted in the 5–7 run total) and “main” runs. In the rest runs, the images were presented at 1 Hz (1 image/sec) for approximately 1.5 minutes, followed by a fixation period of approximately 2 minutes. In the main runs, the images were presented at 2 Hz (0.5 sec intervals), with each run containing 870–871 images and lasting approximately 9 minutes. Each image in both the training and test datasets was presented six times. The presentation order of the categories was randomized across runs.

*MEG cortical current source estimation and preprocessing*

The raw MEG data were preprocessed using Brainstorm software[53]. Environmental noise was first reduced using a high-pass filter at 0.5 Hz and a notch filter at 60 Hz and its harmonics. Independent component analysis (ICA) was then applied to identify and remove cardiac and ocular artifacts. The noise covariance matrix was computed from baseline periods (-500 to 0 ms).

For MEG-MRI coregistration, individual cortical surface models were constructed from T1-weighted MRI anatomical images using FreeSurfer software (Martinos Center Software)[54]. Each subject's three-dimensional facial surface was scanned and aligned with the MRI-derived anatomical facial surface using 100 digitized scalp points (FastSCAN Cobra; Polhemus, Colchester, VT, USA). For source estimation, 15,002 elementary current dipoles were distributed across the cortical surface and oriented perpendicular to the local cortical surface. The forward model was computed using an overlapping sphere head model fitted to the individual cortex tessellation. The inverse problem was then solved using minimum norm estimation (MNE), with the source covariance matrix set to the identity matrix and the regularization parameter $\lambda = 0.1$. The estimated source activities were projected onto the FsAverage template for group analysis. Stimulus onset was marked using analog triggers. All the processes were performed by using Brainstorm.

*MEG ROI identification and selection*

To extract data from each ROI in our MEG recordings, we used the Human Connectome Project Multi-Modal Parcellation 1.0 (HCP-MMP 1.0) atlas[55]. This atlas provides a comprehensive parcellation of the cerebral cortex, dividing each hemisphere into 180 distinct cortical areas (360 areas total). These areas are further organized into 22 larger regions on the basis of anatomical and topographical criteria. The regions consist of adjacent cortical areas that can be viewed completely from one perspective, either on the inflated cortical surface or through flatmap visualization[55]. For our analyses, we focused on the MEG signals extracted from regions within the ventral visual cortex as defined by the HCP-MMP1.0 parcellation.

## ECoG dataset

*Subjects*

In this study, seventeen subjects with normal or corrected-to-normal vision participated in the image presentation tasks (six males; 26.7 ± 11.0 years old; mean ± standard deviation (SD)). All the participants had drug-resistant epilepsy and underwent intracranial electrode implantation as part of their epilepsy treatment (number of subdural electrodes: 64.9 ± 19.4; number of depth electrodes: 6.2 ± 9.3). The subjects were recruited from three university hospitals (Osaka University, Juntendo University, and Nara Medical University). All the participants provided written informed consent after receiving a detailed explanation of the purpose and procedures of the experiment. The study protocol was approved by the institutional ethics committees at each hospital (Osaka University Medical Hospital: Approval No. 14353, UMIN000017900; Juntendo University Hospital: Approval No. 18–164; Nara Medical University Hospital: Approval No. 2098).

*Sample size*

The duration of data collection varied among participants and was dependent on both their clinical treatment schedules and their voluntary participation time. The number of experimental trials was established on the basis of our previous study[12].

*Localization of intracranial electrodes*

The process of localizing intracranial electrodes was performed using presurgical T1-weighted magnetic resonance (MR) images and postsurgical computed tomography (CT) images as follows. Individual cortical surfaces were extracted from MR images and registered to the fsaverage template brain using FreeSurfer[56]. The locations of intracranial electrodes were manually identified on CT images (coregistered to MR images) using BioImage Suite[57]. The identified subdural electrodes were then projected onto individual cortical surfaces using the intracranial electrode visualization toolbox[58]. On the basis of the initial registration, the location of each subdural electrode was mapped to the template brain. For region-based analysis, the electrodes were categorized into 22 brain regions according to the Human Connectome Project parcellation scheme[55]. T1-weighted MRI data of each subject were collected using a 3.0-Tesla SYNAPSE VINCENT scanner (Fujifilm, Tokyo, Japan) located at Osaka University's hospital.

*Stimuli dataset*

Similar to the fMRI and MEG experiments, we used the GOD image dataset, which consists of 1200 training images from 150 categories (8 images per category) and 50 test images from 50 categories (1 image per category). The baseline image dataset consisted of sixty images, including five images each from three categories: faces, landscapes, and words. These images were extracted from the stimulus movies used in a previous study[12]. To create these datasets, all the images were preprocessed by cropping them into squares using the methods outlined in a previous study[15]. The GOD image dataset and images used as the baseline stimuli had no overlap.

*ECoG acquisition*

The subjects viewed visual stimuli while seated either on hospital beds or in chairs facing a computer screen. ECoG signals were acquired using an EEG-1200 system (Nihon Koden,

Tokyo, Japan) at a 10 kHz sampling rate, with reference to the average of two intracranial electrodes. A DATAPixx3 system (VPixx Technologies, Quebec, Canada) monitored the presentation timing of visual stimuli, synchronizing this information with the ECoG recordings.

*Experimental settings*

The image presentation task was conducted over multiple recording sessions (2–4 sessions across 1–3 days), with baseline ECoG recordings acquired at the start of each session to compensate for electrode impedance variations[59]. During all the tasks, the participants maintained fixation on a central point displayed on the screen. Four subjects from the initial cohort were selected for detailed analysis on the basis of two criteria: the presence of electrode implants in the ventral stream visual cortex and above-chance initial samplewise and dimensionwise decoding performance. These selected subjects had 74, 56, 30, and 71 electrodes implanted, respectively.

*Baseline recording task*

To account for electrode impedance variations between recording sessions, a baseline recording task was conducted at the beginning of each session. The task comprised one run, during which baseline dataset images were presented sequentially in random order for 1125 ± 25 ms each, without intervening blank screens.

*Image presentation task*

All the participants participated in the image presentation task, where visuals from the GOD image dataset were shown as stimuli. Each training session consisted of two runs to display all the GOD training images, while each test session included a single run. Within each run, 10 images from the preceding stimulus dataset were presented first in a randomized order, followed by randomly ordered images from the GOD dataset. No blank intervals separated the images, and each image was displayed for approximately 525 ± 25 milliseconds.

*Signal preprocessing and calculation of high-γ features*

For each subject, we performed a visual inspection of the raw data and excluded noisy channels from all subsequent analyses. Common average referencing was then applied to mitigate common noise sources and accentuate local neural activity. ECoG epochs, which were time-locked to stimulus onset and extended 0.5 seconds after the stimulus, were extracted to focus on stimulus-related processing. Power spectral density (PSD) analysis was performed on each epoch using Welch's method with 1024-sample windows, and the high gamma power component (80–150 Hz) was extracted by summing the power within this frequency range for each channel. To complete the preprocessing pipeline, we concatenated the data corresponding to electrodes that were placed in the ventral visual stream of patients as the final ECoG data for the subsequent decoding analyses.

*Autoencoder framework*

The autoencoder consists of two fully connected layers with ReLU activation functions. The number of dimensions in the autoencoder's latent space was set to half the number of dimensions of the original vectors. For each subject, brain region and semantic space type, a separate autoencoder was trained. When training the autoencoder for a particular subject, we used the averaged brain RSMs of all the other subjects. After we finished the training process, we passed all the original semantic vectors to the trained model and used the intermediate features of the resulting trained autoencoder as the brain-aligned features.

The RSM matrices were created from the brain activity pattern or the autoencoder's latent space by calculating the pairwise cosine similarity of each of the two data points. During the training process, we used the difference between the upper triangle of each of the RSM matrices to constrain the autoencoder to make representations more brain-like.

**Neural decoding of visual stimuli**

We performed brain decoding by constructing linear regression models to predict semantic vectors from brain activity patterns. To predict each unit of semantic vectors, a separate set of linear regression models was trained. Prior to regression analysis, we performed voxel selection via a method similar to that used by Horikawa and Kamitani (2017), and the brain activity patterns were Z-normalized.

More formally, given that $x = \{x_1, x_2, \ldots, x_n\}^T$ represents the activity of $n$ neural activity data points (i.e., voxels in the fMRI data, source-estimated neural activity patterns from MEG sensors, and neural amplitude recorded from each channel in each second in the ECoG) from the region of interest, the regression function can be represented as follows:

$$y(x) = \sum_{i=1}^{n} w_i x_i + w_0$$

where $x_i$ is a scalar value specifying the amplitude of the brain data point $i$, $w_i$ is the weight of voxel $i$ and $w_0$ is the bias.

For each subject, semantic space type and brain region, we trained a separate set of linear regression functions as decoders. When the fMRI data of a particular subject were decoded to the brain-aligned semantic spaces, we used the brain-aligned space in which that subject was not used to create. When the MEG data of a particular subject were decoded, we used the averaged brain-aligned semantic spaces of all the fMRI subjects.

## Identification analysis

For the identification analysis, the predicted vector was identified among a large set of candidate vectors. First, we prepared one random image from 1000 randomly selected classes of the ImageNet dataset. Then, for each semantic space (i.e., the GloVe- and CLIP-based pretrained feature vectors or the GloVe- and CLIP-based brain-aligned vectors for different values of $\alpha$), we calculated the corresponding semantic vectors of all the images that had been randomly selected from ImageNet. If we could not obtain the GloVe embeddings of a category, that category was excluded from all analyses. After obtaining the brain-aligned vectors of all the ImageNet categories, we input the original GloVe/CLIP-pretrained feature vectors to the corresponding trained autoencoder and obtained the corresponding brain-aligned vectors. Then, for each category in the GOD dataset, we calculated the Pearson correlation coefficient between the true and predicted vectors and between the predicted vector and all other candidate vectors and assigned the identification accuracy as the percentage of candidate categories, in which their correlation with the test predicted vector is lower than the correlation of the true and predicted vectors. The chance-level identification accuracy was determined by randomly shuffling the true feature vectors and calculating the identification accuracy for the shuffled vectors, following the same procedure as for the unshuffled data.

## Statistical tests

In the decoding analyses, we evaluated the performance of the brain decoders using the Pearson correlation coefficients between the predicted and true feature vectors as well as between the predicted and shuffled true feature vectors. We then applied Fisher's z-transform to the correlations of each case to stabilize variance, followed by one-sided *t*-tests for each feature space type and neuroimaging modality. Similarly, in the identification analysis, we performed a one-sided *t*-test between the identification results of shuffled data and unshuffled data.

To compare the decoding and identification accuracy means among the original feature vectors and brain-aligned feature vectors, we applied one-way analysis of variance (ANOVA) followed by Tukey's honestly significant difference (HSD) post hoc test. Prior to each *t*-test and ANOVA, we assessed the normality of the data via the Shapiro–Wilk test.

To calculate the significant differences in *F* value distributions among the different types of CLIP-based feature vectors (original vs. brain-aligned), we applied the two-sided Wilcoxon rank-sum test between each pairwise combination of brain-aligned feature vectors.

**Acknowledgments**

We acknowledge the use of open-source code from the Kamitani Lab. Specifically, we used the Brain Decoding Toolbox (BDPy; https://github.com/KamitaniLab/bdpy) for neuroimaging data processing and analysis, and adapted decoding algorithms from the Generic Object Decoding repository (https://github.com/KamitaniLab/GenericObjectDecoding; Horikawa & Kamitani, 2017). We thank the Kamitani Lab for making these resources publicly available. We also thank all the subjects for their participation. This research was supported by the Japan Science and Technology Agency (JST) Moonshot R&D (JPMJMS2012), the JST Core Research for Evolutional Science and Technology (CREST) (JPMJCR18A5), the JST AIP Acceleration Research (JPMJCR24U2), and the Japan Society for the Promotion of Science (JSPS) Grants-in-Aid for Scientific Research (KAKENHI) (JP26560467 and JP20H05705).

**Author contributions**

S.V. and T.Y. conceptualized the project. S.V. was responsible for the theory. S.V., R.F. and T.Y. were responsible for the methodology. S.V. undertook the analysis and investigation. R. F. and H. Y. were responsible for the MEG and ECoG experiments. S.V. was responsible for data preprocessing and curation. S.V. wrote the original draft and created the figures. S.V. and T.Y edited the final version of the article. S.O., N.T., H.M.K., H.S., Y.I., H.S., M.N., H.K. and K.T. performed the neurosurgery of ECoG experiments.

**Competing interests**

The authors declare that they have no competing interests.

## Data Availability

The datasets supporting the findings of this study include fMRI, MEG, and ECoG data. The fMRI dataset used in this study is available at[15]. The MEG and ECoG datasets are available from the corresponding author upon reasonable request and will be provided under a formal data-sharing agreement.

## Code Availability

The code used for data analysis in this study will become publicly available on our GitHub repository (https://github.com/yanagisawa-lab) at the time of publication. For inquiries about the code before it is uploaded, please contact the corresponding author.

## Figures and Figure legends

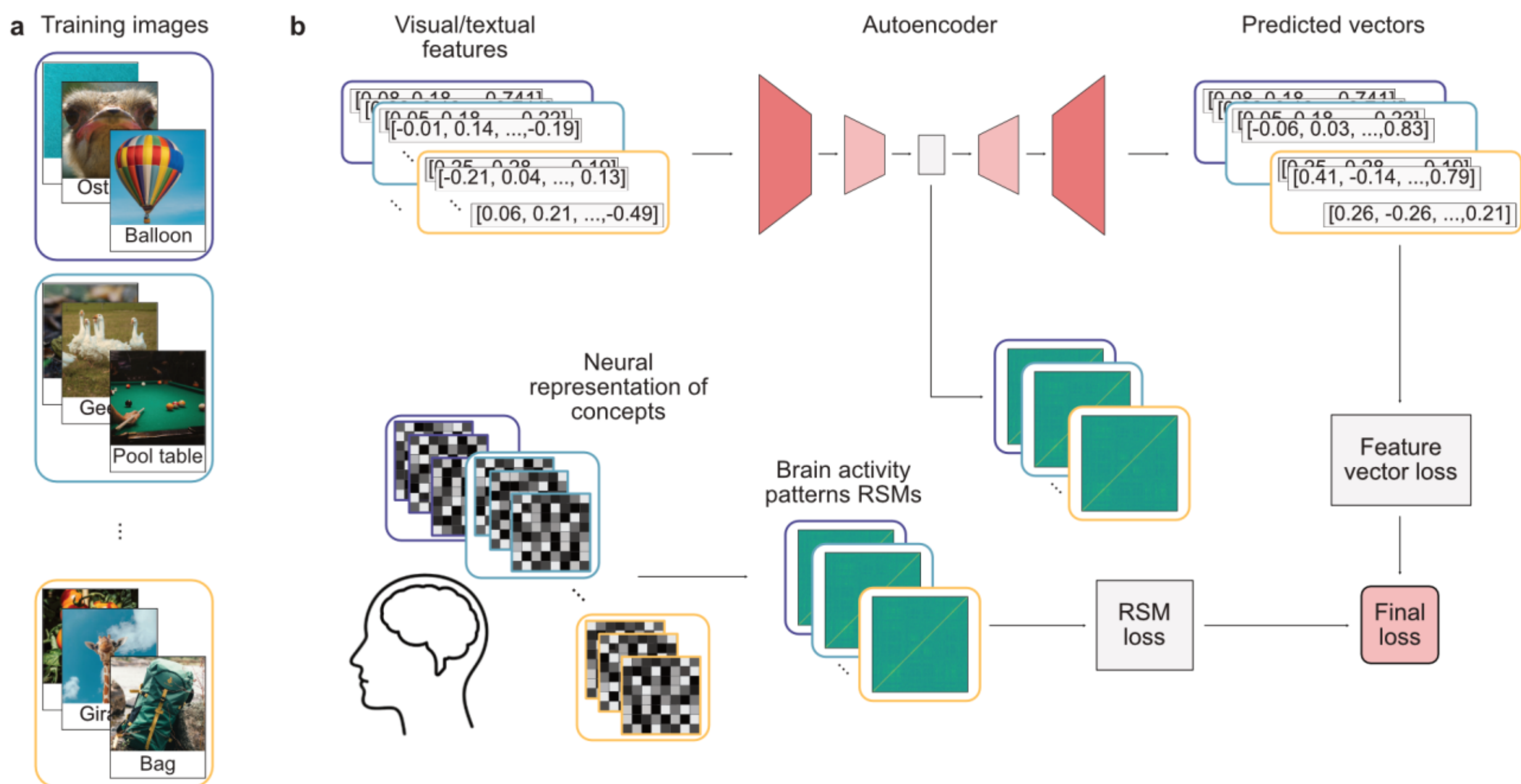

**Fig. 1. Examples of the visual stimuli and workflow of the brain-aligning framework. a** Sample images from the GOD dataset. The rectangles represent the sample image batches used for training. **b** Brain-aligning framework. First, pretrained visual or textual features are extracted. Then, an autoencoder is trained to reconstruct these features while aligning the representational similarity matrix (RSM) of its latent space with the RSM of corresponding brain activity patterns.

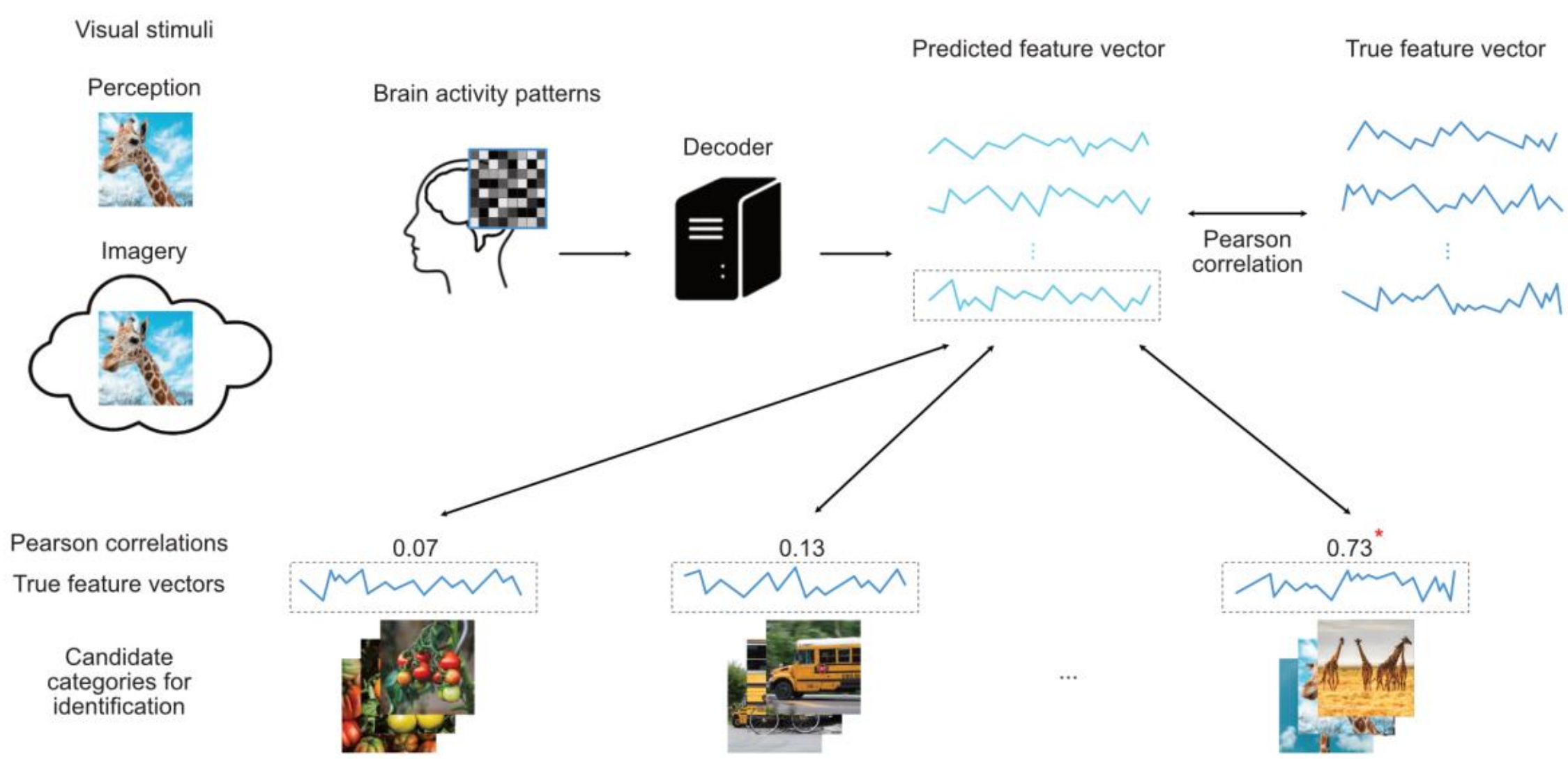

**Fig. 2. Brain decoding and stimulus identification procedure.** Brain decoders are trained to map neural activity patterns from visual stimuli (perception or imagery) to their corresponding semantic feature vectors. For stimulus identification, the predicted semantic vector is compared against a large set of candidate stimulus categories using Pearson correlation coefficients. The stimulus with the highest correlation to the predicted vector is identified as the decoded stimulus. The identification accuracy is calculated as the percentage of candidate categories with lower correlations than the true target stimulus.

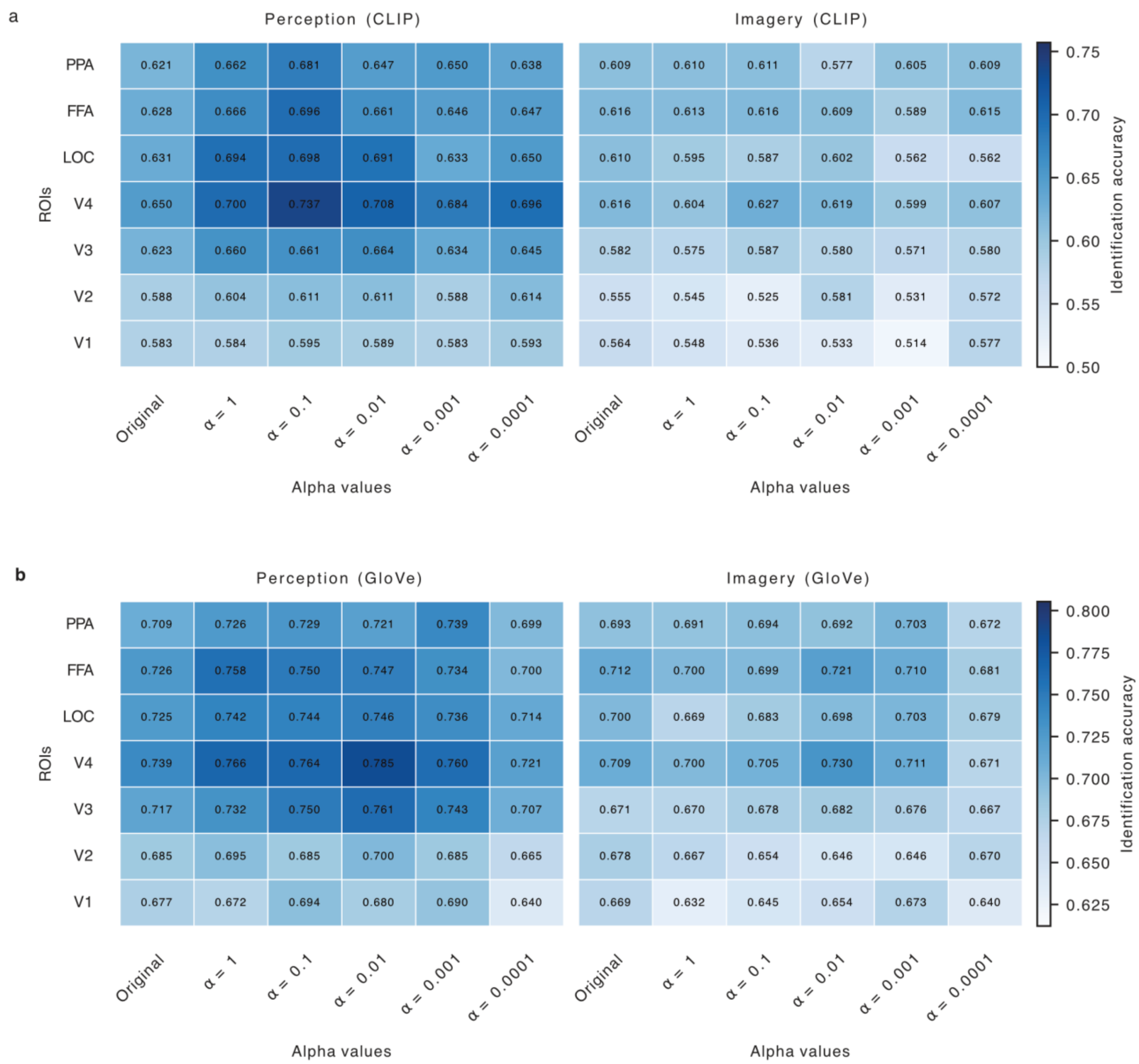

**Fig. 3. fMRI identification accuracy patterns using brain-aligned semantic vectors across visual cortex regions and alignment parameters.** Heatmaps showing zero-shot identification accuracy for visual stimulus decoding using brain-aligned semantic vectors derived from different regions of interest (ROIs) in the visual cortex. The results are shown for **a** CLIP-based and **b** GloVe-based semantic vectors under both perception and imagery conditions. The rows represent different visual cortex ROIs (V1–V4: primary and secondary visual areas; LOC: lateral occipital complex; FFA: fusiform face area; PPA: parahippocampal place area). The columns

represent different brain-alignment parameters (α), where "Original" indicates unaligned pretrained vectors, $\alpha = 1$ indicates reconstruction without brain alignment, and decreasing α values (0.1 to 0.0001) indicate increasing degrees of brain alignment. The color intensity reflects identification accuracy, with warmer colors indicating better performance. The optimal brain-alignment parameters varied by semantic vector type and brain region, with V4-derived brain-aligned vectors showing particularly strong performance for both CLIP ($\alpha = 0.1$) and GloVe ($\alpha = 0.01$) across perception and imagery conditions. $n = 5$ subjects from the GOD dataset.

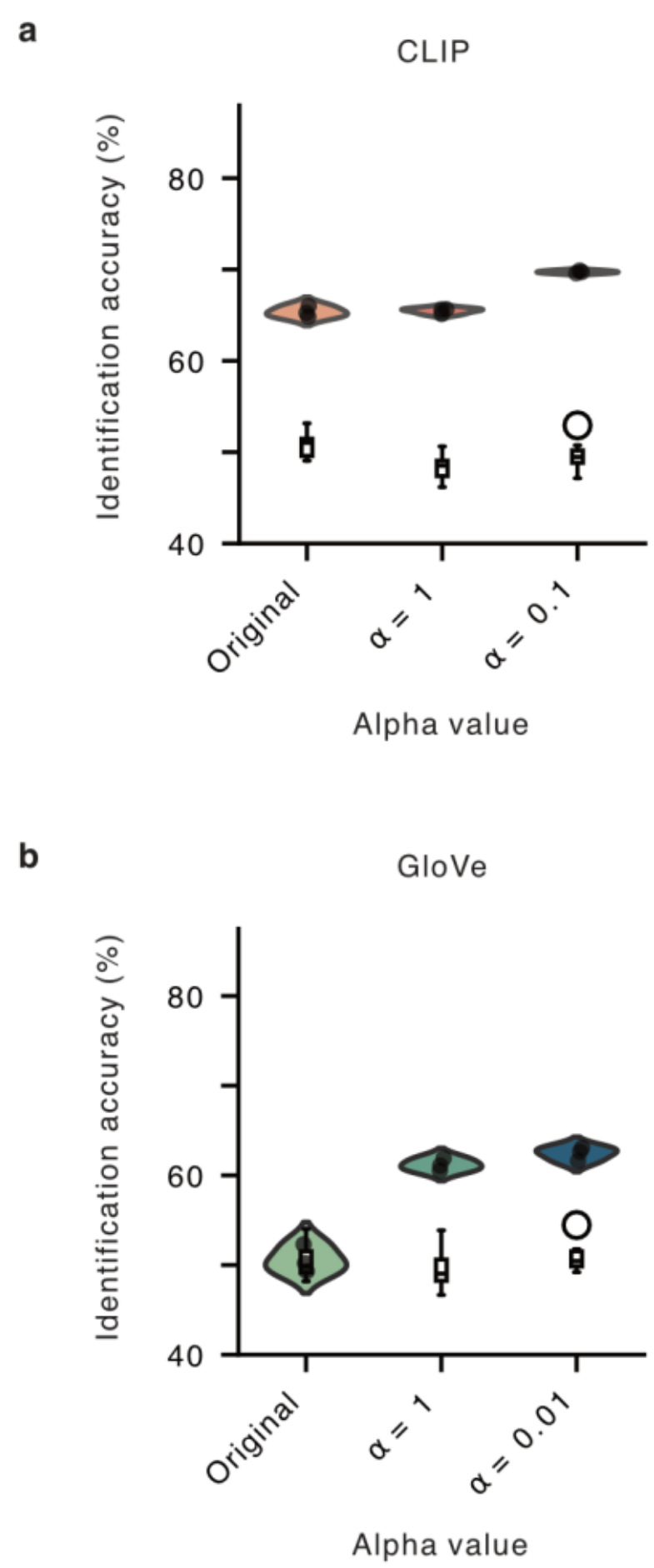

**Fig. 4. Cross-modal generalization: MEG identification accuracy using the fMRI-derived brain-aligned semantic vectors.** Violin plots showing zero-shot identification accuracy for visual stimulus decoding in MEG data using brain-aligned semantic vectors derived from the fMRI V4 region. **a** CLIP-based and **b** GloVe-based semantic vectors with optimal α parameters determined from the fMRI results. The individual data points from 3 subjects are overlaid on violin plots. The white circles indicate shuffled control data, demonstrating chance-level performance.

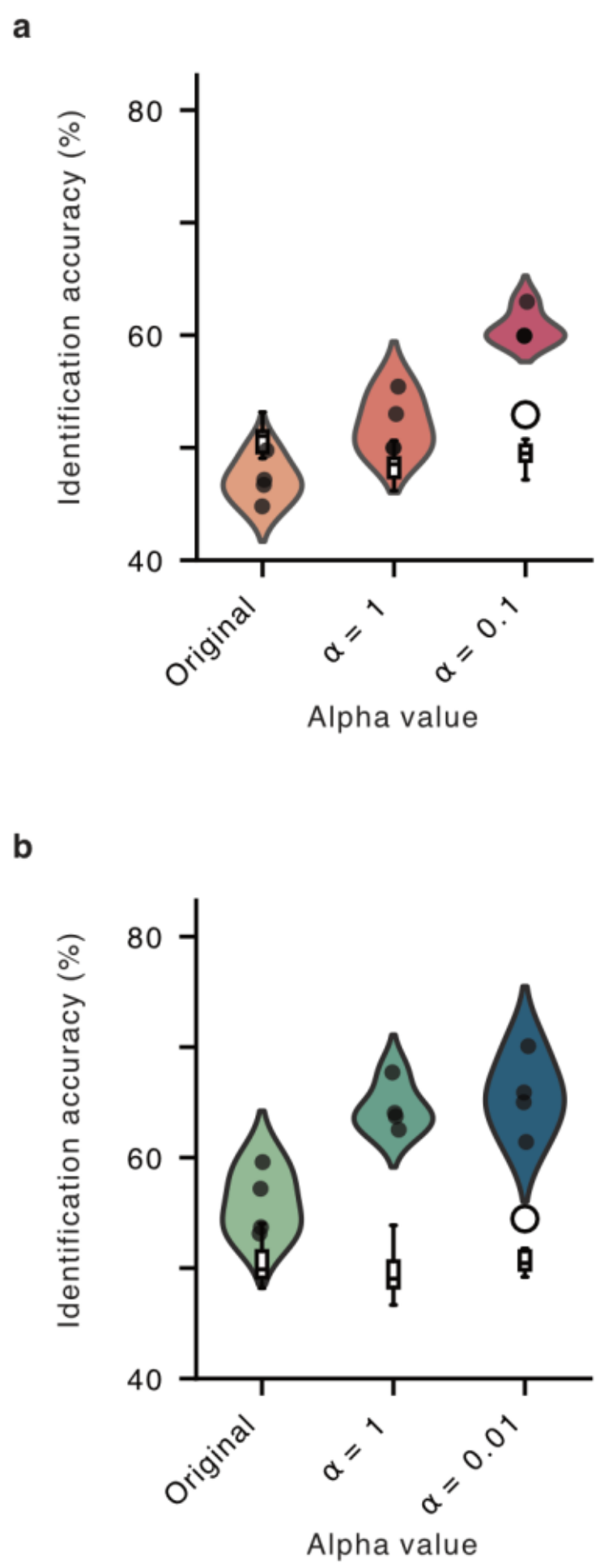

**Fig. 5. Cross-modal generalization: ECoG identification accuracy patterns using the fMRI-derived brain-aligned semantic vectors.** Violin plots showing zero-shot identification accuracy for visual stimulus decoding in ECoG data using brain-aligned semantic vectors derived from the fMRI V4 region. **a** CLIP-based and **b** GloVe-based semantic vectors with optimal α parameters determined from the fMRI results. The individual data points from 4 subjects (E1–E4) are overlaid on violin plots. The white circles indicate shuffled control data, demonstrating chance-level performance.

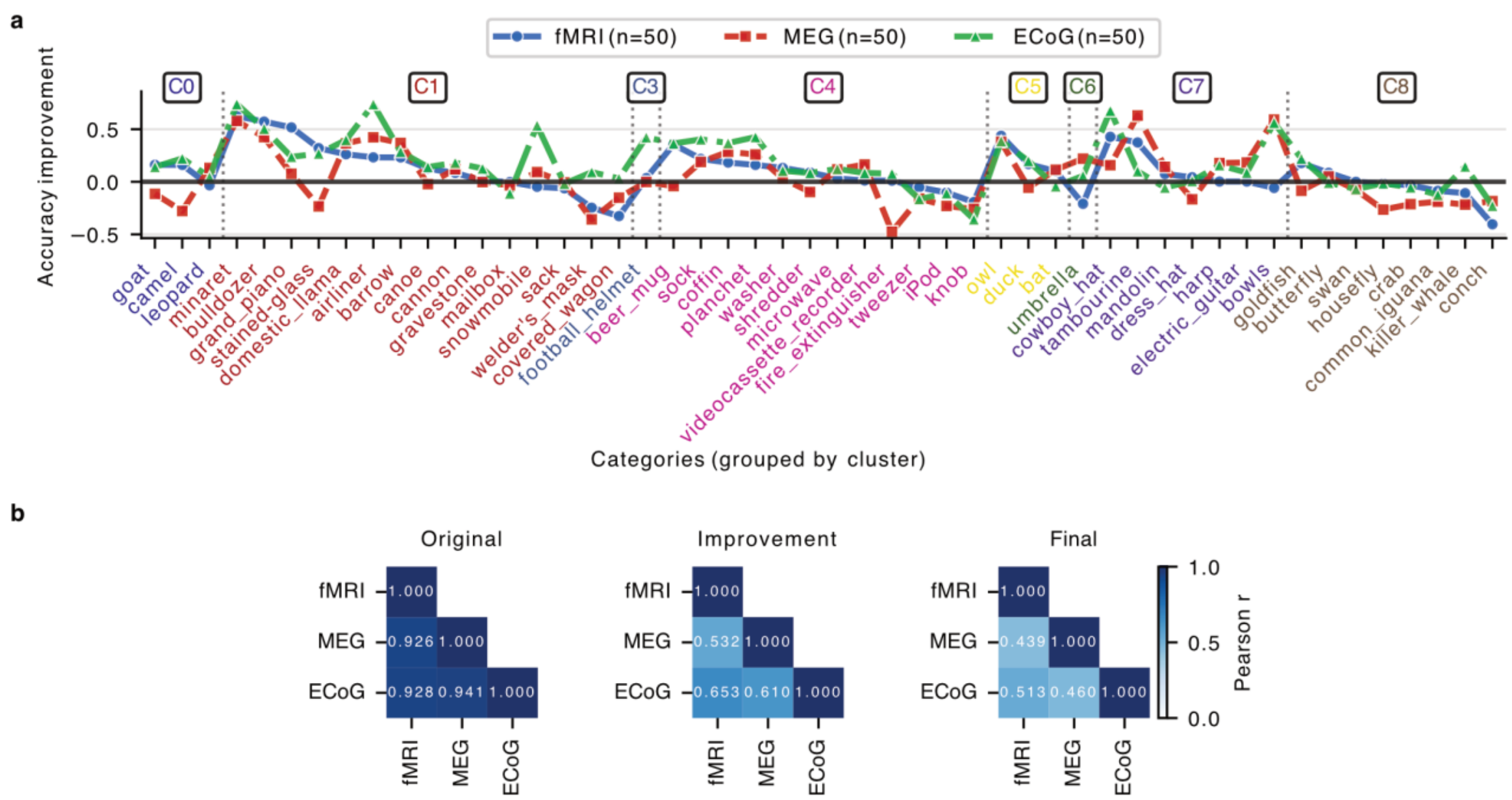

**Fig. 6. Consistency of identification accuracy improvement between the optimal CLIP-based brain-aligned vectors and original CLIP-based vectors across all three modalities.** **a** Category-specific accuracy improvements following brain alignment of CLIP vectors across the fMRI (blue circles), MEG (red squares), and ECoG (green triangles) modalities. The brain-aligned vectors were trained on the fMRI V4 region data, with the optimal α parameters determined from the fMRI V4 region identification analysis and applied across all the modalities. The categories are grouped by semantic clusters (C0–C8) derived from whole ImageNet clustering analysis (Supplementary Figure 6). Each modality tested the same n=50 categories. The y-axis shows the accuracy improvement relative to the original CLIP vectors. The dotted vertical lines separate the cluster boundaries. The consistent patterns in identification accuracy improvements across modalities demonstrate robust cross-modal generalization of V4 region brain-alignment benefits. **b** Pearson correlation matrices comparing the accuracy patterns across neuroimaging modalities for original performance (left), improvement (center), and final performance (right). All the results use brain-aligned vectors trained on fMRI V4 region data with optimal α parameters derived from fMRI V4 region identification analysis. The values

indicate correlation coefficients between modality pairs. The high correlations in the accuracy improvement matrix (center) demonstrate that categories benefiting from V4 brain alignment in fMRI consistently benefit across the MEG and ECoG modalities, supporting the robustness and generalizability of the brain alignment approach.

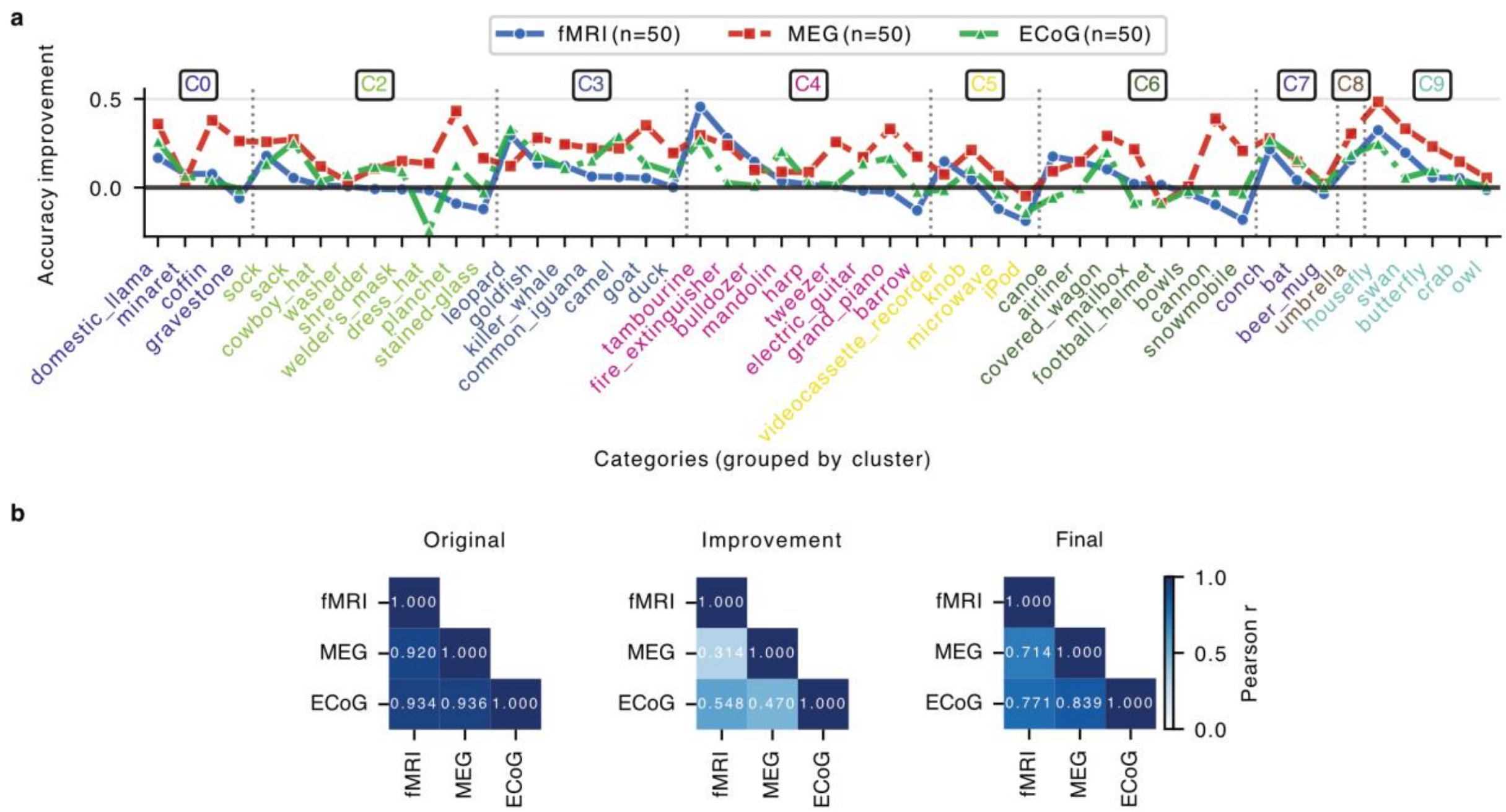

**Fig. 7. Consistency of identification accuracy improvement between the optimal GloVe-based brain-aligned vectors and original GloVe-based vectors across all three modalities. a** Category-specific accuracy improvements following brain alignment of GloVe vectors across the fMRI (blue circles), MEG (red squares), and ECoG (green triangles) modalities. The brain-aligned vectors were trained on the fMRI V4 region data, with the optimal α parameters determined from the fMRI V4 region identification analysis and applied across all the modalities. The categories are grouped by semantic clusters (C0–C9) derived from whole ImageNet clustering analysis (Supplementary Figure 7). Each modality tested the same n=50 categories. The y-axis shows the accuracy improvements relative to the original GloVe vectors. The dotted vertical lines separate the cluster boundaries. The consistent patterns in identification accuracy improvements across modalities demonstrate robust cross-modal generalization of V4 region brain-alignment benefits. **b** Pearson correlation matrices comparing the accuracy patterns across neuroimaging modalities for original performance (left), improvement (center), and final performance (right). All the results use brain-aligned vectors trained on fMRI V4 region data

with optimal α parameters derived from fMRI V4 region identification analysis. The values indicate correlation coefficients between modality pairs. The high correlations in the accuracy improvement matrix (center) demonstrate that categories benefiting from V4 brain alignment in fMRI consistently benefit across the MEG and ECoG modalities, supporting the robustness and generalizability of the brain alignment approach.

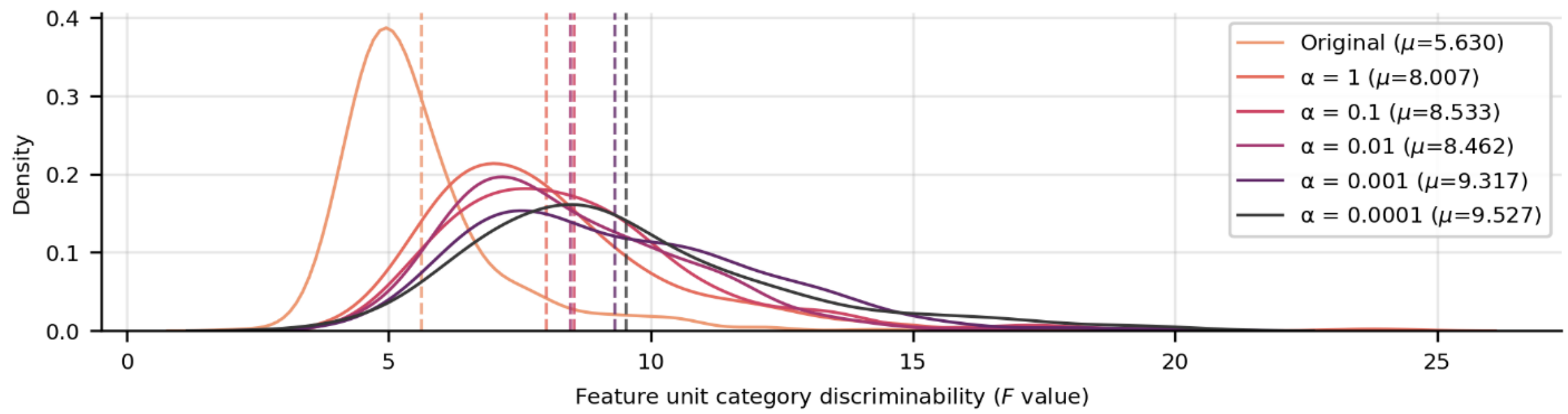

**Fig. 8. Brain alignment progressively enhances feature unit category discriminability.** Kernel density distributions of the *F* statistics measuring category discriminability across feature units for the original CLIP features and brain-aligned variants. Stronger brain alignment (smaller α) systematically shifts the distributions rightward, indicating improved discriminability. The dashed vertical lines indicate the mean *F* statistic for each condition.

**Supplementary information for:**

# Brain-aligning of semantic vectors improves neural decoding of visual stimuli

Shirin Vafaei[1], Ryohei Fukuma[1,2], *Takufumi Yanagisawa[1,2,3], Huixiang Yang[2], Satoru Oshino[1,3], Naoki Tani[1,3], Hui Ming Khoo[1,3], Hidenori Sugano[4], Yasushi Iimura[4], Hiroharu Suzuki[4], Madoka Nakajima[4], Kentaro Tamura[5,6], Haruhiko Kishima[1]

[1]Department of Neurosurgery, Graduate School of Medicine, Osaka University, Suita, Japan

[2]Institute for Advanced Co-Creation Studies, Osaka University, Suita, Japan

[3]Osaka University Hospital Epilepsy Center, Suita, Japan

[4]Department of Neurosurgery, Juntendo University, Tokyo, Japan

[5]Department of Neurosurgery, Nara Medical University, Kashihara, Japan

[6]Department of Neurosurgery, National Hospital Organization, Nara Medical Center, Nara, Japan

***Corresponding author**

Takufumi Yanagisawa

Address: 2-2 Yamadaoka, Suita, Osaka 565-0871, Japan

Tel.: +81-6-6879-3652

Fax: +81-6-6879-3659

E-mail: tyanagisawa@nsurg.med.osaka-u.ac.jp

## Supplementary Notes

To evaluate decoding accuracy comprehensively, we also calculated the samplewise correlation accuracy, which was calculated as the average Pearson correlation between the true and predicted vectors across all samples and subjects and is a global measure of the ability of the model to reconstruct entire brain activity patterns[1–4]. Additionally, we analyzed the dimensionwise correlation, which was computed as the average Pearson correlation across feature units, and assessed how well the model reconstructed each individual dimension of the learned representational space[5–9].

When the decoding accuracy was measured with samplewise and dimensionwise metrics, we observed that the decoding accuracy depended on the degree of brain-aligning and that higher accuracy was achieved with lower alpha values (Supplementary Figs. 15 and 16).

Similar to the fMRI results, the MEG and ECoG samplewise and dimensionwise decoding accuracies depended on the degree of brain alignment, and higher accuracy was achieved with lower alpha values (see Supplementary Figs. 17–20).

To further validate the importance of the semantic structure in the original feature space for effective brain alignment, we conducted additional experiments using one-hot encoded category representations. As shown in Supplementary Figure 21, when the autoencoder was trained on discrete one-hot vectors lacking inherent semantic relationships, the identification accuracy remained near chance levels across all α values, despite successful reconstruction of the binary representations. These results confirm that brain alignment requires semantically meaningful input representations to be effective—the autoencoder cannot create a semantic structure from purely categorical labels but rather relies on the rich semantic relationships already present in pretrained feature spaces such as CLIP and GloVe. Additionally, our analysis of training dynamics revealed that the amount of brain data that were used for autoencoder training significantly influenced convergence speed (Supplementary Figure 22). Networks trained with larger brain sample sizes (1200 vs. 300 samples) demonstrated faster and more stable convergence across both CLIP- and GloVe-based vectors, suggesting that sufficient neural data are crucial for capturing robust brain-semantic alignments. Together, these findings highlight two critical requirements for successful brain alignment: semantically structured input representations and adequate neural training data to capture the statistical properties of brain activity patterns.

## Supplementary Figures

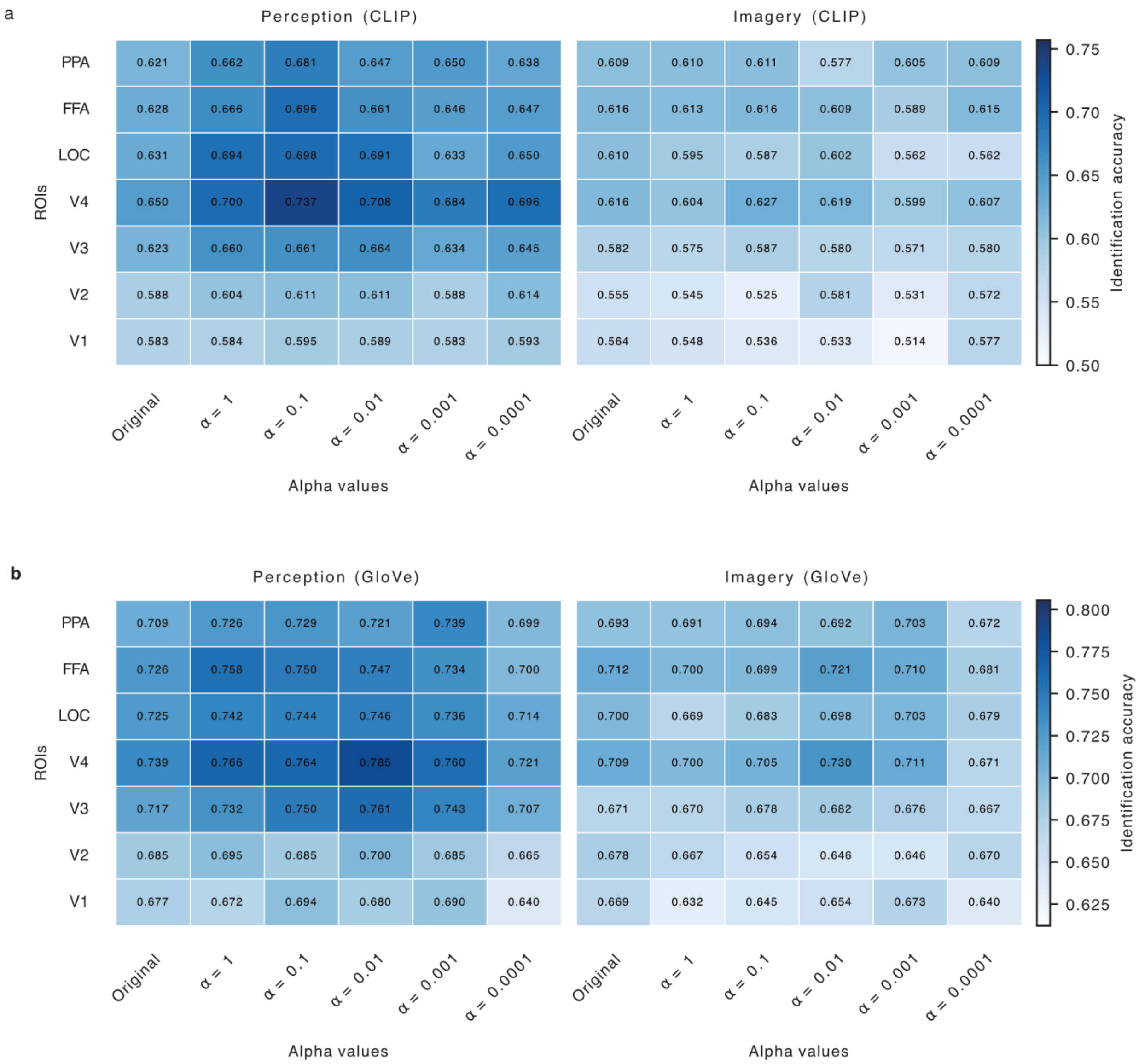

**Supplementary Fig. 1 | fMRI identification accuracy patterns using the brain-aligned semantic vectors across visual cortex regions and alignment parameters.** Heatmaps showing the zero-shot identification accuracy for visual stimulus decoding using brain-aligned semantic vectors derived from different regions of interest (ROIs) in the visual cortex. The results are shown for the (a) CLIP-based and (b) GloVe-based semantic vectors under both the perception and imagery conditions. The rows represent different visual cortex ROIs (V1–V4: primary and

secondary visual areas; LOC: lateral occipital complex; FFA: fusiform face area; PPA: parahippocampal place area). The columns represent different brain-alignment parameters (α), where "Original" indicates unaligned pretrained vectors, α = 1 indicates reconstruction without brain alignment, and decreasing α values (0.1 to 0.0001) indicate increasing degrees of brain alignment. The color intensity reflects the identification accuracy, with warmer colors indicating better performance. The optimal brain-alignment parameters varied by semantic vector type and brain region, with V4-derived brain-aligned vectors showing particularly strong performance for both CLIP (α = 0.1) and GloVe (α = 0.01) across perception and imagery conditions. n = 5 subjects from the GOD dataset. We first evaluated the performance of category identification relative to chance. Specifically, we compared the identification results obtained from the original data with those obtained from shuffled data. The shuffled accuracy was obtained by correlating the predicted vectors with both the shuffled true vector and all the other candidate vectors. The shuffled identification accuracy was then defined as the percentage of candidate categories whose correlation with the predicted vector was lower than the correlation between the shuffled true and predicted vectors. The real identification accuracies were significantly greater than the corresponding shuffled accuracies for 42/42 of the CLIP perception data, 35/42 of the CLIP imagery, 42/42 of the GloVe perception and 42/42 of the GloVe imagery (one-sided *t* test, $p < 0.05$, see Supplementary Tables 1–4 for the exact values). Two-way ANOVA revealed significant main effects of ROI across all feature types in both perception and imagery, and it also revealed significant main effects of alpha in perception data across all feature types: CLIP perception (ROI: $F_{(6,168)} = 22.67$, $p < 0.001$; Alpha: $F_{(5,168)} = 6.27$, $p < 0.001$), CLIP imagery (ROI: $F_{(6,168)} = 4.73$, $p < 0.001$; Alpha: $F_{(5,168)} = 0.59$, $p = 0.709$), GloVe perception (ROI: $F_{(6,168)} = 42.20$, $p < 0.001$; Alpha: $F_{(5,168)} = 14.76$, $p < 0.001$) and GloVe imagery (ROI: $F_{(6,168)} = 3.81$, *p*

= 0.001; Alpha: $F_{(5,168)} = 0.83$, $p = 0.528$). No significant ROI × alpha interactions were detected (all $p > 0.05$). The higher-order visual areas (V4, LOC, FFA, and PPA) consistently showed greater identification accuracies than the early visual areas did (V1, V2, and V3), with brain-aligned vectors demonstrating enhanced zero-shot identification performance. For the CLIP-based vectors, $\alpha = 0.1$ derived from V4 achieved the highest identification accuracy for both the perception and imagery data, whereas for the GloVe-based vectors, $\alpha = 0.01$ derived from V4 regions showed optimal results across both conditions. These results demonstrate that the best identification accuracy was optimized by aligning the V4 activity patterns with the brain-aligned vectors, with the optimal alignment parameter ($\alpha$) depending on the original semantic space.

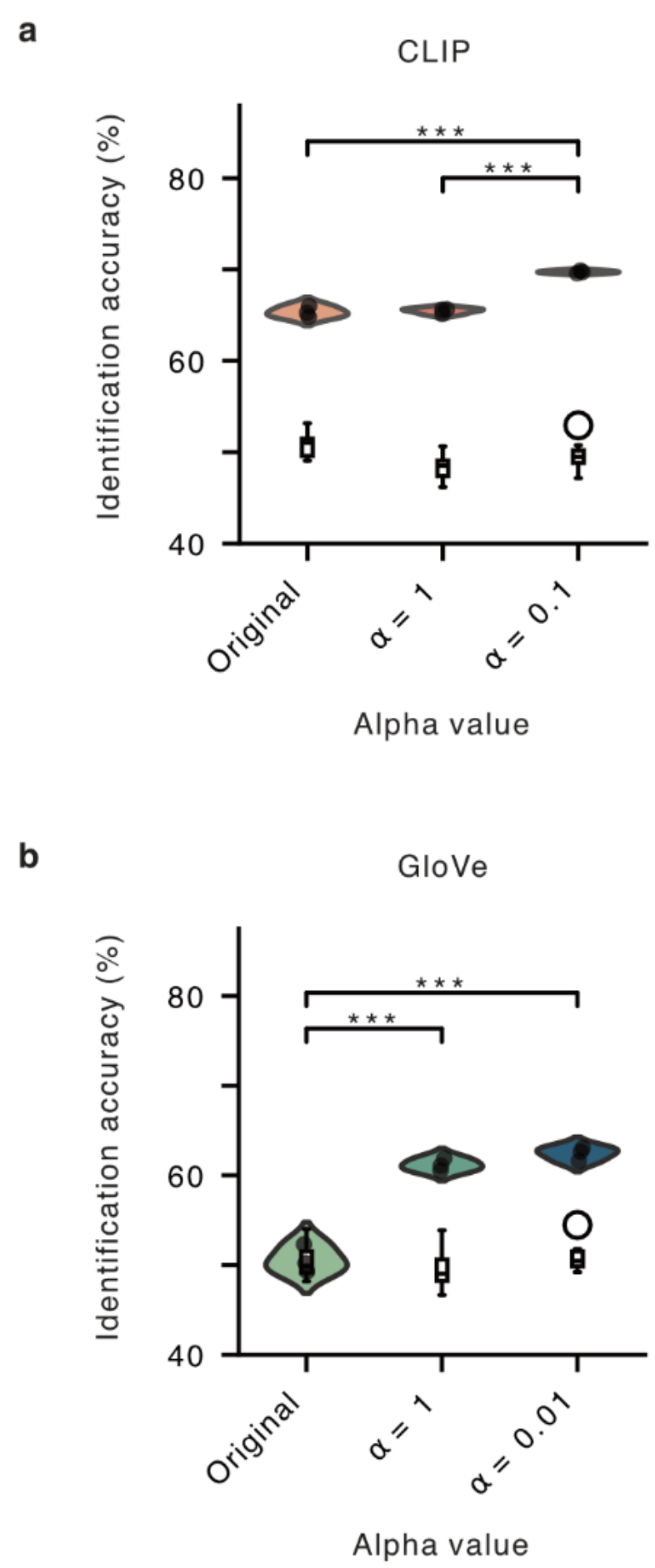

**Supplementary Fig. 2 | Cross-modal generalization: MEG identification accuracy using fMRI-derived brain-aligned semantic vectors.** Violin plots showing zero-shot identification accuracy for visual stimulus decoding in the MEG data using the brain-aligned semantic vectors derived from the fMRI V4 region. **a.** CLIP-based and **b.** GloVe-based semantic vectors with the optimal α parameters determined from the fMRI results. The individual data points from 3 subjects are overlaid on the violin plots. The white circles indicate the shuffled control data, demonstrating chance-level performance. The significance bars indicate the post hoc comparisons (*** $p < 0.001$). The error bars represent the standard error of the mean across subjects. For the CLIP-based vectors, significant main effects of α were observed ($F_{(2, 6)}$ =

116.51, $p < 0.001$, one-way ANOVA). α = 0.1 vectors significantly outperformed both the original vectors ($p < 0.001$, Tukey's HSD test) and α = 1 vectors ($p < 0.001$, Tukey's HSD test), whereas no significant difference was observed between the original and α = 1 vectors ($p$ = 0.938, Tukey's HSD test). For the GloVe-based vectors, significant main effects of α were also found ($F_{(2, 6)} = 110.85$, $p < 0.001$, one-way ANOVA). The brain-aligned vectors with α = 0.01 and α = 1 significantly outperformed the original vectors ($p < 0.001$, Tukey's HSD test). However, no significant difference was found between the two brain-aligned conditions (α = 0.01 vs. α = 1, $p = 0.377$, Tukey's HSD test). These findings indicate that the optimal brain alignment parameters identified in the fMRI data (α = 0.1 for CLIP, α = 0.01 for GloVe) are successfully generalized to MEG neural decoding, demonstrating robust cross-modal transferability of the brain-aligned semantic representations. For the subject-by-subject results, see Supplementary Tables 11–12.

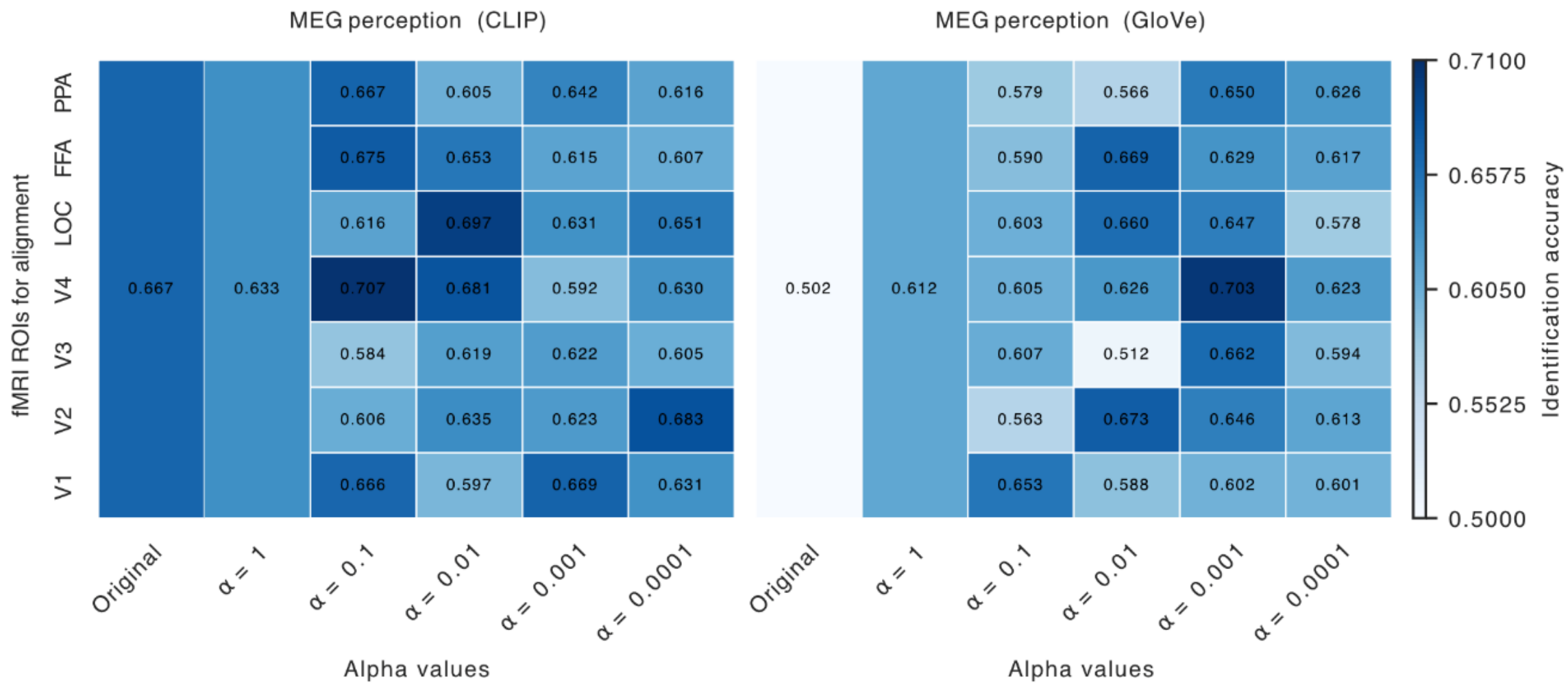

**Supplementary Fig. 3 | Comprehensive MEG identification results using the brain-aligned semantic vectors derived from different fMRI regions.** Heatmaps showing the zero-shot identification accuracy for visual stimulus decoding in the MEG data using the brain-aligned semantic vectors derived from different fMRI regions of interest. The results are shown for the CLIP-based (left) and GloVe-based (right) semantic vectors. The rows represent the different visual cortex ROIs used for brain alignment training (V1–V4: primary and secondary visual areas; LOC: lateral occipital complex; FFA: fusiform face area; PPA: parahippocampal place area). The columns represent different brain-alignment parameters α, where "Original" indicates unaligned pretrained vectors, α = 1 indicates reconstruction without brain alignment, and decreasing α values indicate increasing degrees of brain alignment. The color intensity reflects the identification accuracy, with warmer colors indicating better performance. n = 3 MEG subjects. Cell-by-cell one-sided *t* tests compared the real identification accuracy against the shuffled controls, revealing that 38/42 CLIP conditions (90.5%) and 28/42 GloVe conditions (66.7%) achieved significantly above-chance performance ($p < 0.05$; see Supplementary Tables 9–10 for complete results).

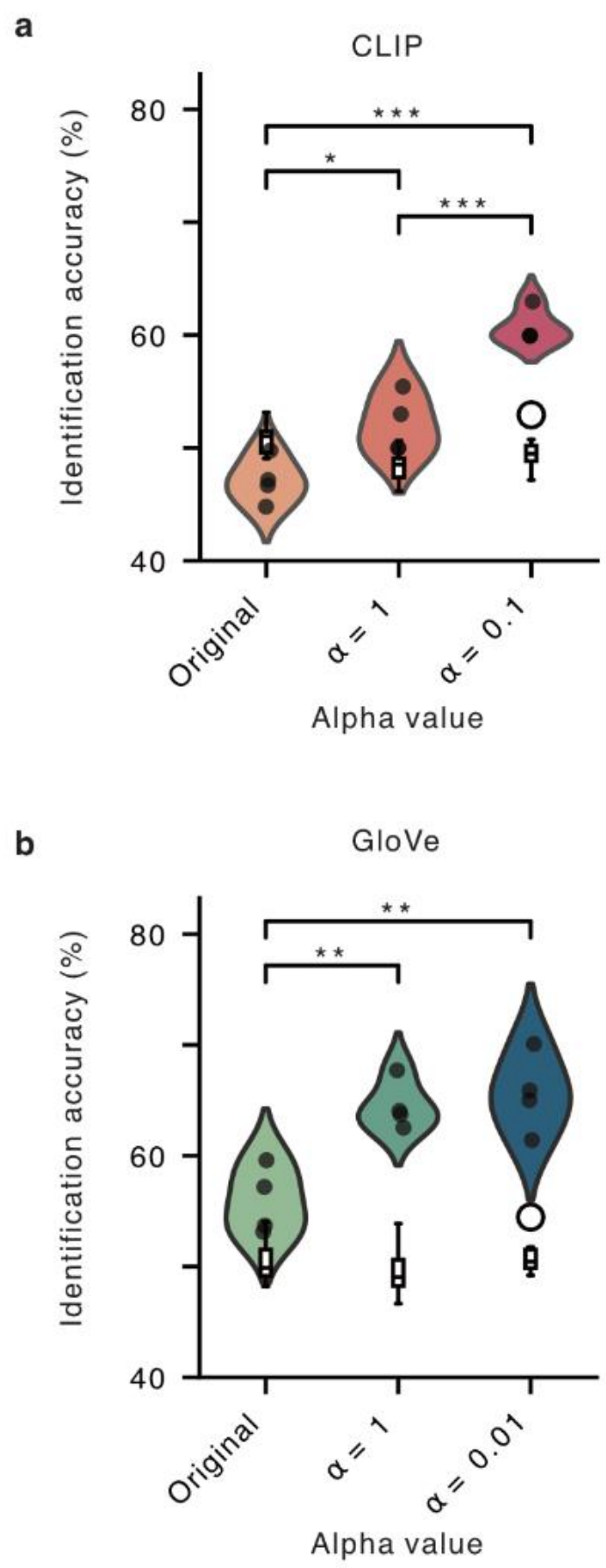

**Supplementary Fig. 4 | Cross-modal generalization: ECoG identification accuracy patterns using the fMRI-derived brain-aligned semantic vectors.** Violin plots showing the zero-shot identification accuracy for visual stimulus decoding in the ECoG data using the brain-aligned semantic vectors derived from the fMRI V4 region. **a** CLIP-based and **b** GloVe-based semantic vectors with the optimal α parameters determined from the fMRI results. The individual data points from 4 subjects (E1–E4) are overlaid on the violin plots. The white circles indicate the shuffled control data, demonstrating chance-level performance. The significance bars indicate the post hoc comparisons (* $p < 0.05$, ** $p < 0.01$, *** $p < 0.001$). Next, we examined the differences in the identification accuracy between the original vectors, vectors with α = 1, and

the optimal brain-aligned vectors from the fMRI results. For the CLIP-based vectors, significant main effects of α were observed ($F(2, 9) = 42.2549$, $p < 0.001$, one-way ANOVA). The α = 0.1 vectors significantly outperformed both the original vectors ($p < 0.001$, Tukey's HSD test; Fig 4a) and the α = 1 vectors ($p = 0.0007$, Tukey's HSD test; Fig 4a), while the α = 1 vectors also significantly outperformed the original vectors ($p = 0.0213$, Tukey's HSD test, Fig 4a). For the GloVe-based vectors, significant main effects of α were also found ($F(2, 9) = 12.5537$, $p = 0.0025$, one-way ANOVA). The brain-aligned vectors with α = 0.01 and α = 1 both significantly outperformed the original vectors ($p = 0.0034$ and $p = 0.0072$, respectively, Tukey's HSD test, Fig 4b). However, no significant difference was found between the two brain-aligned conditions (α = 0.01 vs. α = 1, $p > 0.05$, Tukey's HSD test, Fig 4b). These findings indicate that the optimal brain-alignment parameters identified in the fMRI data (α = 0.1 for CLIP, α = 0.01 for GloVe) are successfully generalized to ECoG neural decoding, demonstrating robust cross-modal transferability of the brain-aligned semantic representations. For the subject-by-subject results, see Supplementary Tables 15–16.

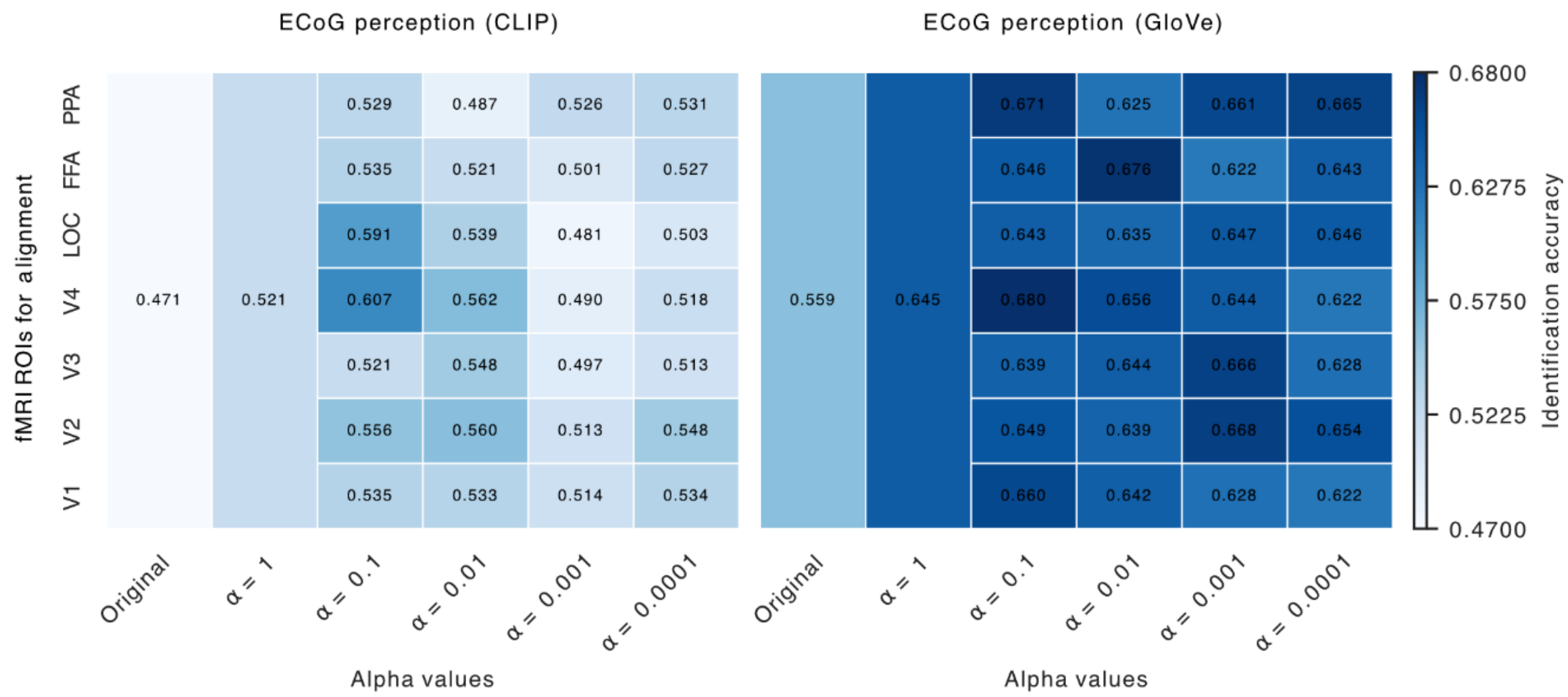

**Supplementary Fig. 5 | ECoG identification accuracy patterns using the brain-aligned semantic vectors across the fMRI-derived visual cortex alignment parameters.** Heatmaps showing the zero-shot identification accuracy for visual stimulus decoding using the brain-aligned semantic vectors in the ECoG data. The results are shown for the CLIP-based (left) and GloVe-based semantic vectors (right) under the perception conditions. The rows represent the different fMRI-derived visual cortex ROIs used for brain alignment (V1–V4: primary and secondary visual areas; LOC: lateral occipital complex; FFA: fusiform face area; PPA: parahippocampal place area). The columns represent the different brain-alignment parameters (α), where "Original" indicates the unaligned pretrained vectors, α = 1 indicates reconstruction without brain alignment, and decreasing α values (0.1 to 0.0001) indicate increasing degrees of brain alignment. The color intensity reflects the identification accuracy, with warmer colors indicating better performance. Compared with the CLIP-based vectors, the GloVe-based vectors demonstrated substantially higher identification accuracies across all conditions. For the CLIP-based vectors, V4-derived brain alignment with α = 0.1 achieved the highest identification accuracy, whereas for the GloVe-based vectors, V4-derived alignment with α = 0.1 also yielded

the optimal results. Statistical significance testing revealed that the CLIP identification accuracies were significantly greater than chance-level for 25/42 conditions (one-sided *t* test, $p < 0.05$, see Supplementary Table 13), whereas the GloVe identification accuracies were significantly greater than baseline for 35/42 conditions, with most showing high significance ($p < 0.001$, see Supplementary Table 14). The brain-aligned vectors consistently outperformed the original unaligned vectors, with the higher-order visual areas (V4, LOC, FFA, and PPA) generally showing superior alignment effectiveness compared with the early visual areas (V1, V2, V3). These results demonstrate that fMRI-derived brain alignment can enhance ECoG-based visual stimulus identification, with the optimal alignment parameters and brain regions varying between semantic vector types.

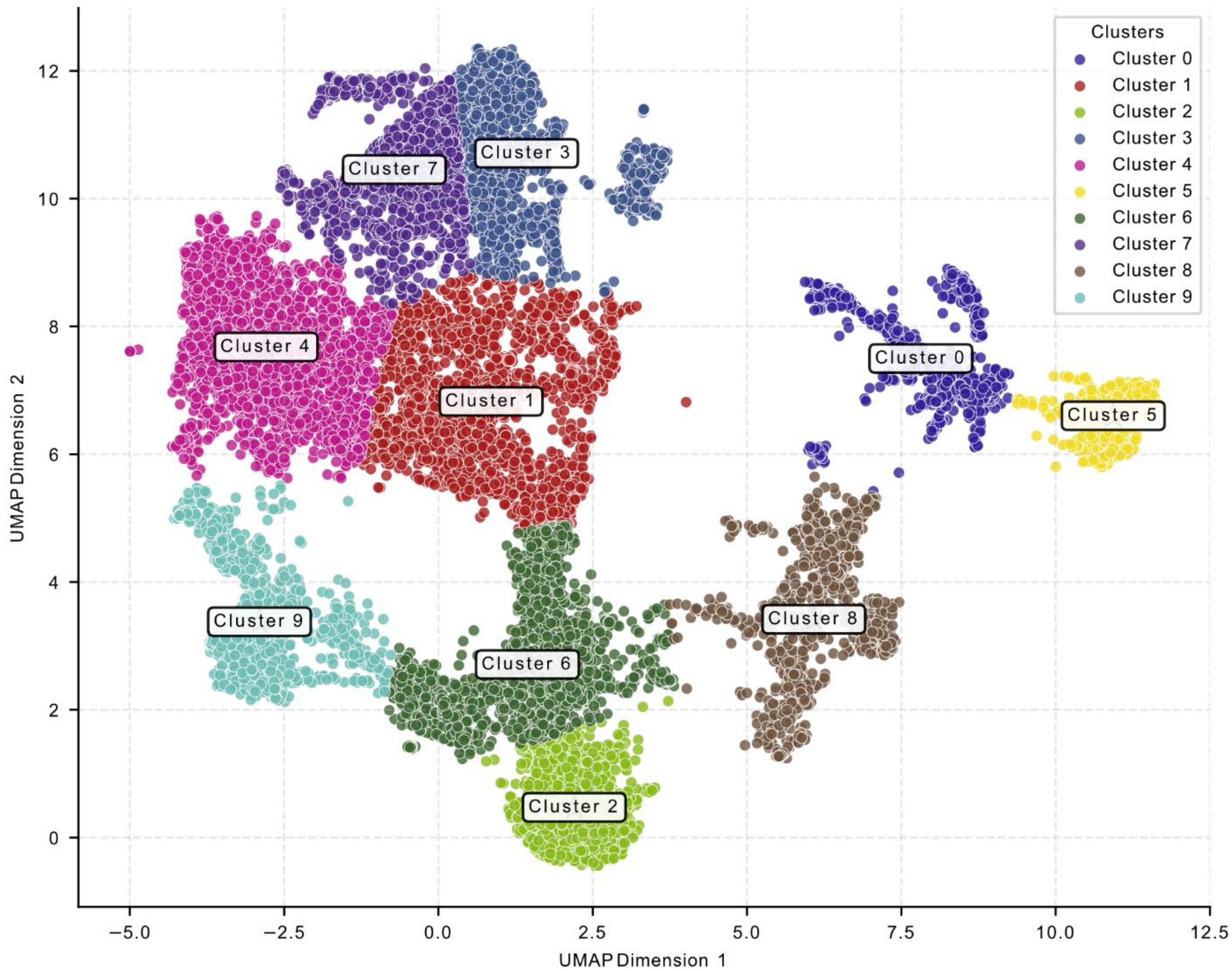

**Supplementary Fig. 6 | CLIP-based clustering of the ImageNet categories.** Feature vectors were extracted for each ImageNet category using the CLIP image encoder model, followed by dimensionality reduction using uniform manifold approximation and projection (UMAP)[10] with a cosine distance metric, which preserves both the local neighborhood structure and the global topology while projecting the high-dimensional features into 2D space. K-means clustering was then applied to partition the categories into 10 distinct clusters. Each point represents an individual ImageNet category, colored according to the cluster assignment.

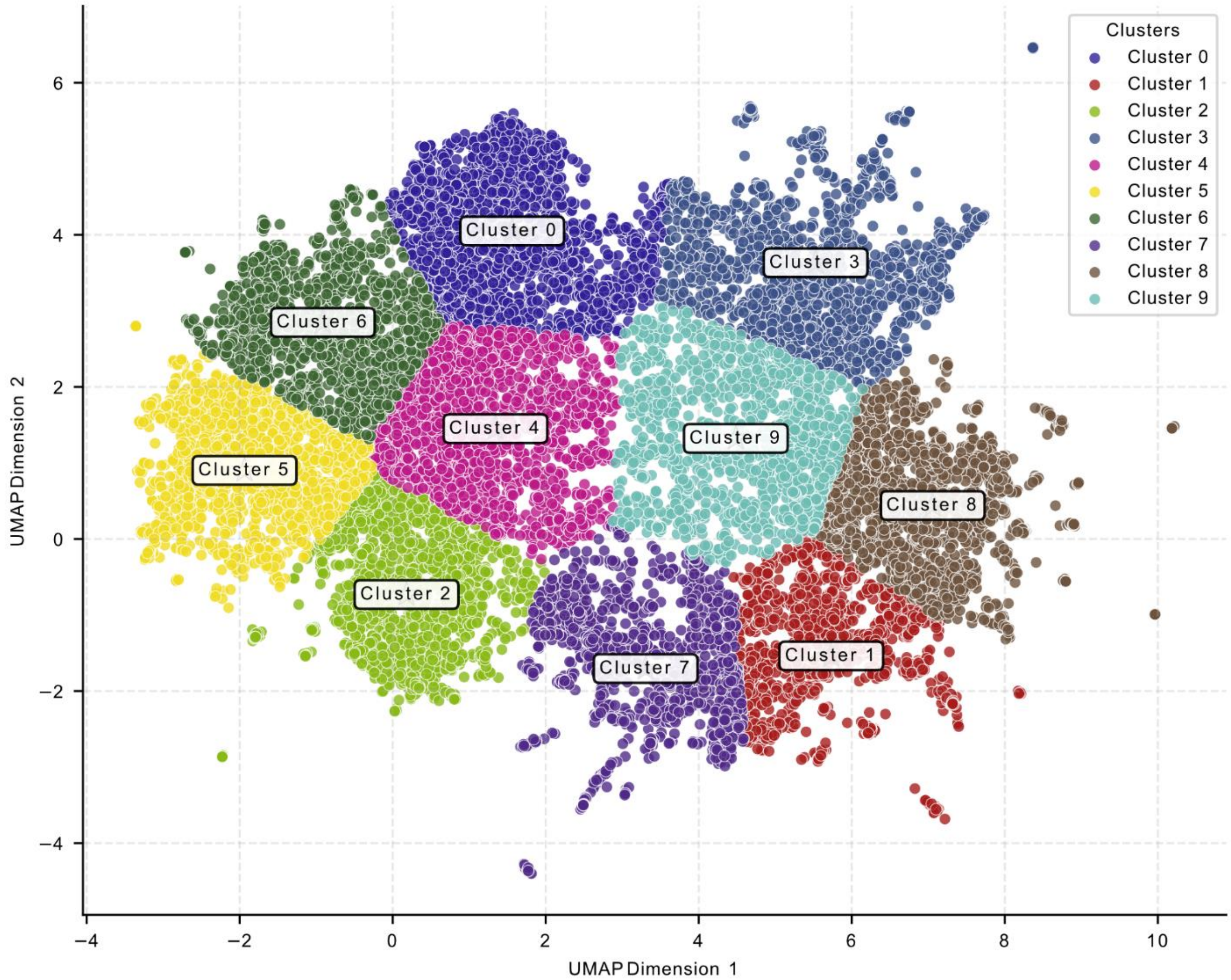

**Supplementary Fig. 7 | GloVe-based clustering of the ImageNet categories.** Feature vectors were extracted for each ImageNet category using pretrained GloVe embeddings, followed by dimensionality reduction using UMAP with a cosine distance metric to preserve the manifold structure of the semantic space while reducing the dimensionality from the original embedding space to 2D coordinates. K-means clustering was then applied to partition the categories into 10 distinct clusters. Each point represents an individual ImageNet category, colored according to the cluster assignment.

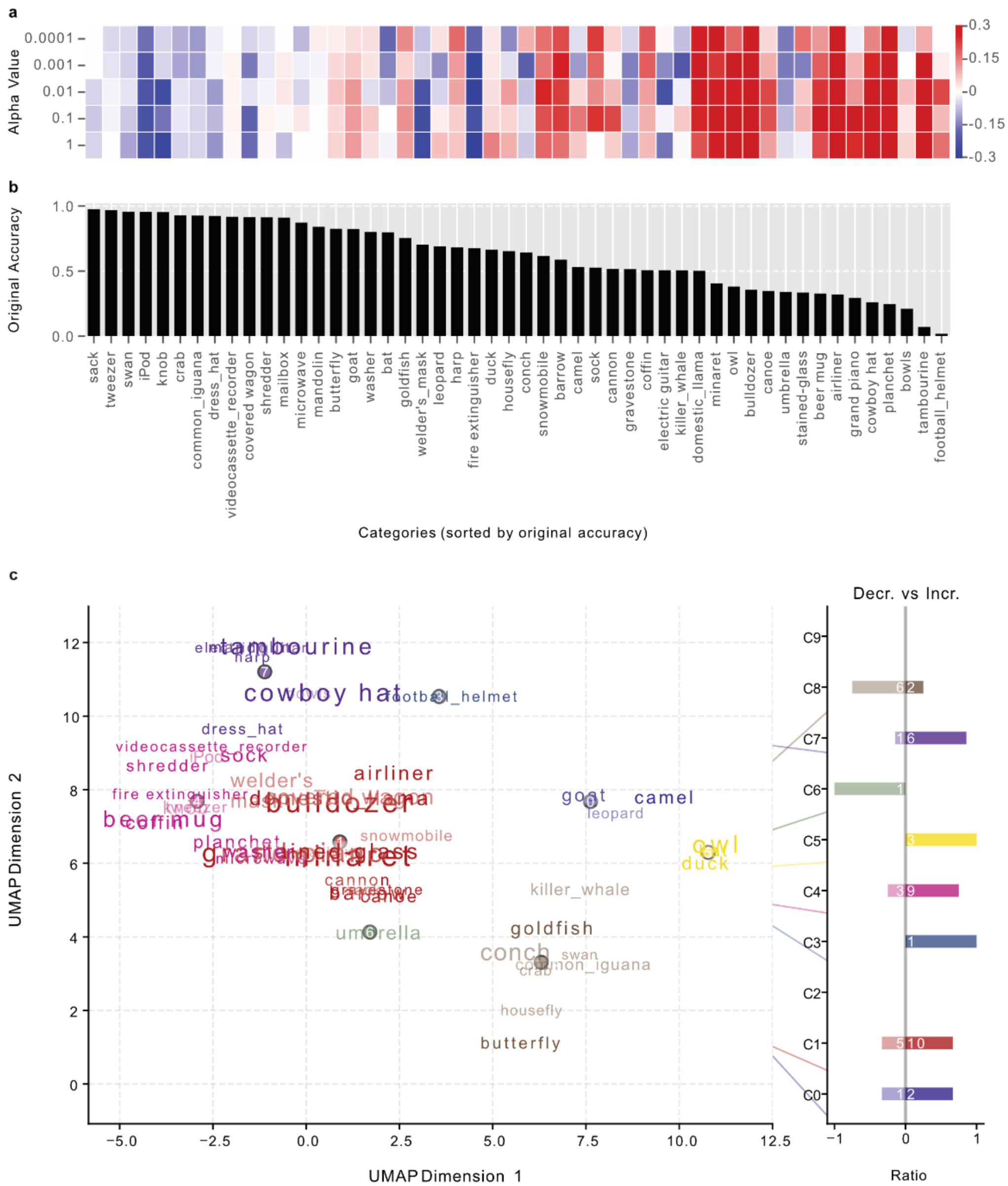

**Supplementary Fig. 8 | Category-specific enhancement patterns of the brain-aligned CLIP vectors in fMRI identification. a.** Heatmap showing the relative change in the identification accuracy across different brain-alignment strengths (α parameters) for each ImageNet category. The categories are ordered identically to those in panel b (high to low original accuracy). The color scale represents the accuracy change relative to the identification accuracy of the original CLIP vectors, with blue indicating decreased performance and red indicating improved

performance. Lower α values represent stronger incorporation of brain activity patterns from the fMRI V4 region during fine-tuning. **b.** Baseline identification accuracy using the original CLIP vectors across the test dataset categories, sorted from highest (left) to lowest (right) performing categories. The accuracy represents the zero-shot identification performance on the held-out test data. **c.** UMAP visualization of the category-specific improvements following brain-alignment optimization. The categories are positioned according to their location in the whole ImageNet clustering analysis (Supplementary Figure 3). The word size corresponds to the magnitude of accuracy change for the best-performing α value of the V4 brain region compared to original CLIP features. The transparent text indicates decreased accuracy, whereas the bold text indicates improved accuracy. The brain-aligned CLIP features show preferential improvements for object categories (e.g., airliner, bulldozer) compared with biological entities, suggesting that CLIP's visual-semantic representations benefit from brain alignment for artifact recognition tasks.

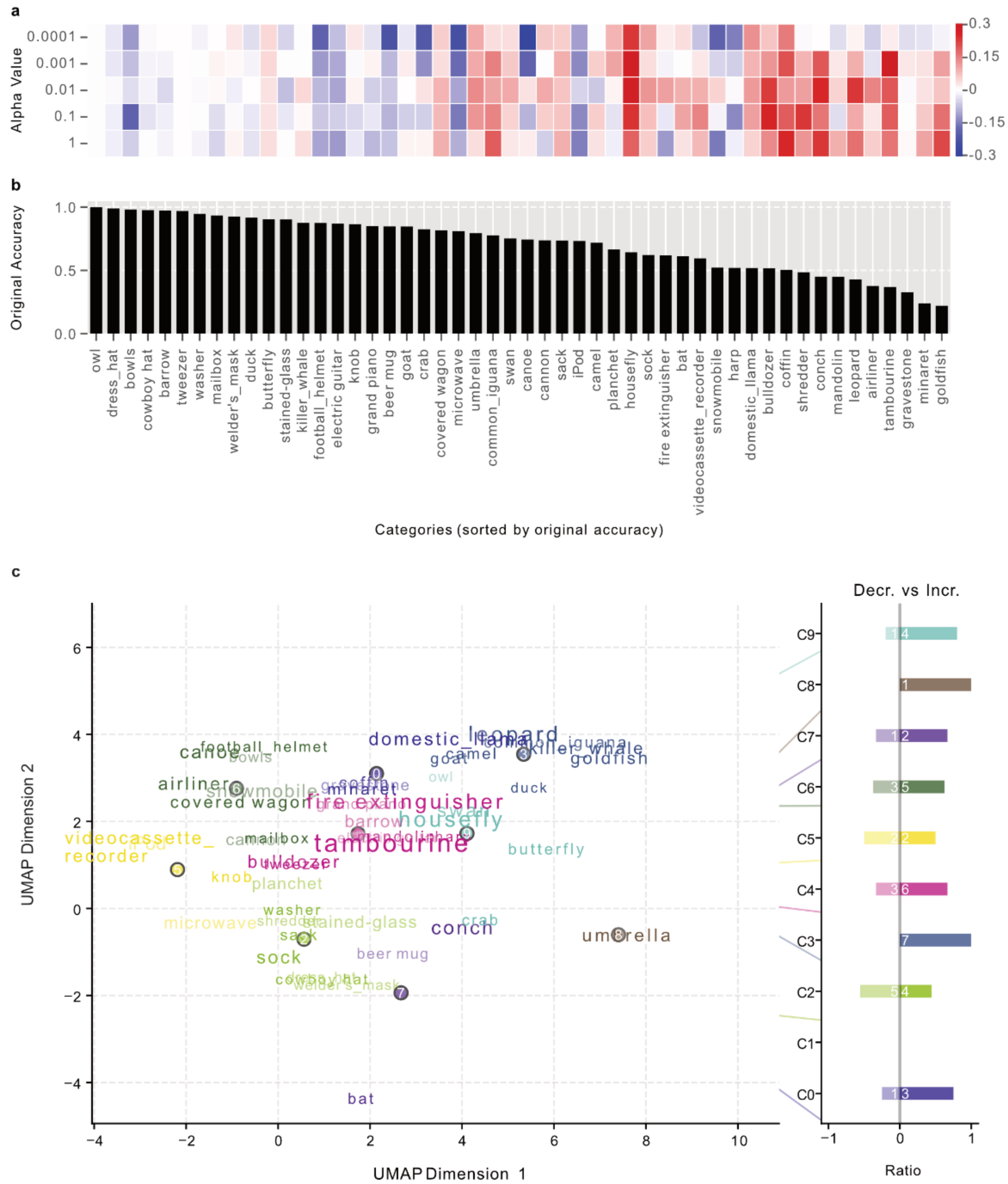

**Supplementary Fig. 9 | Category-specific enhancement patterns of the brain-aligned GloVe features in fMRI identification. a.** Heatmap displaying relative changes in identification accuracy across brain-alignment parameters (α values) for each test dataset category. The categories are arranged in descending order of the original GloVe identification performance (matching panel b order). The color gradient indicates the performance change relative to the baseline GloVe embeddings, with stronger brain incorporation (lower α) showing category-

specific enhancement patterns. **b.** Baseline identification accuracy using original GloVe word embeddings for test dataset categories, ranked from highest to lowest performance. The accuracy reflects the zero-shot identification capability on the independent fMRI test data. **c.** UMAP projection showing differential category improvements after brain-alignment training. The spatial positioning reflects the whole ImageNet semantic clustering structure (Supplementary Figure 4). The text size represents the magnitude of the accuracy improvement or degradation from the optimal α of the V4 brain region compared to original GloVe vectors. The transparent labels indicate performance decreases, while the bold labels indicate accuracy gains. The brain-aligned GloVe embeddings demonstrate preferential enhancement for animal categories (e.g., owl, duck) over inanimate objects, indicating that semantic word representations benefit from neural alignment specifically for biological concept recognition.

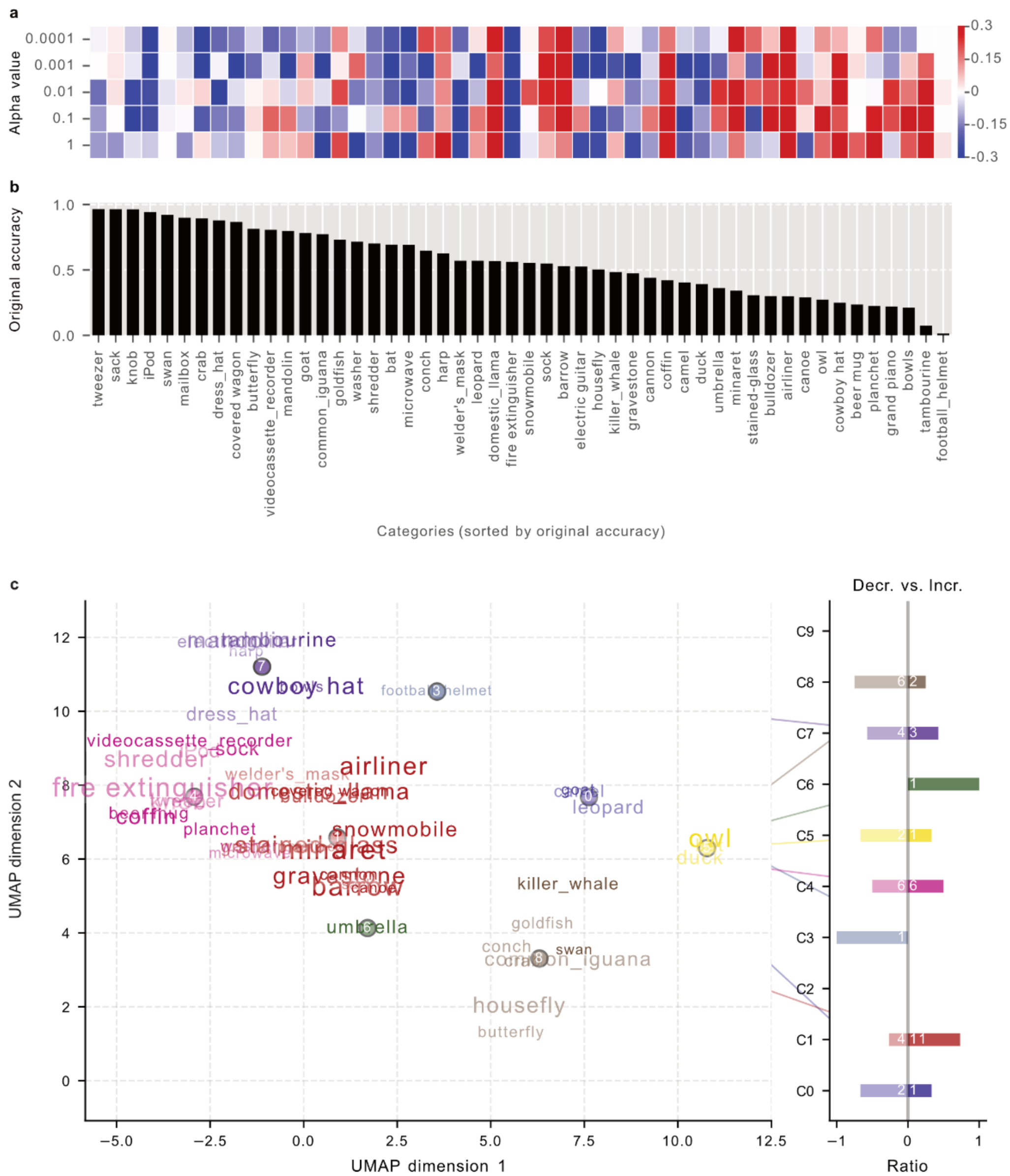

**Supplementary Fig. 10 | Category-specific enhancement patterns of the brain-aligned CLIP vectors in MEG identification. a.** Heatmap showing the relative change in the identification accuracy across different brain-alignment strengths (α parameters) for each test category from ImageNet. The categories are ordered identically to those in panel b (high to low original accuracy). The color scale represents the accuracy change relative to the original CLIP vectors, with blue indicating decreased performance and red indicating improved performance. **b.**

Baseline identification accuracy using the original CLIP vectors across 50 test categories from ImageNet, sorted from highest (left) to lowest (right) performing categories. The accuracy represents the zero-shot identification performance on held-out MEG data. **c.** UMAP visualization of the category-specific improvements following brain-alignment optimization. The categories are positioned according to their location in the whole ImageNet clustering analysis (Supplementary Figure 3). The word size corresponds to the magnitude of the accuracy change for the optimal α value (determined from the fMRI V4 brain region) compared to original CLIP vectors. The transparent text indicates decreased accuracy, whereas the bold text indicates improved accuracy. The results are from the identification accuracies of the optimal α value that was determined from the fMRI V4 region results and applied to the MEG data. The brain-aligned CLIP vectors show preferential improvements for object categories compared with biological entities in the MEG data, which is consistent with the fMRI findings.

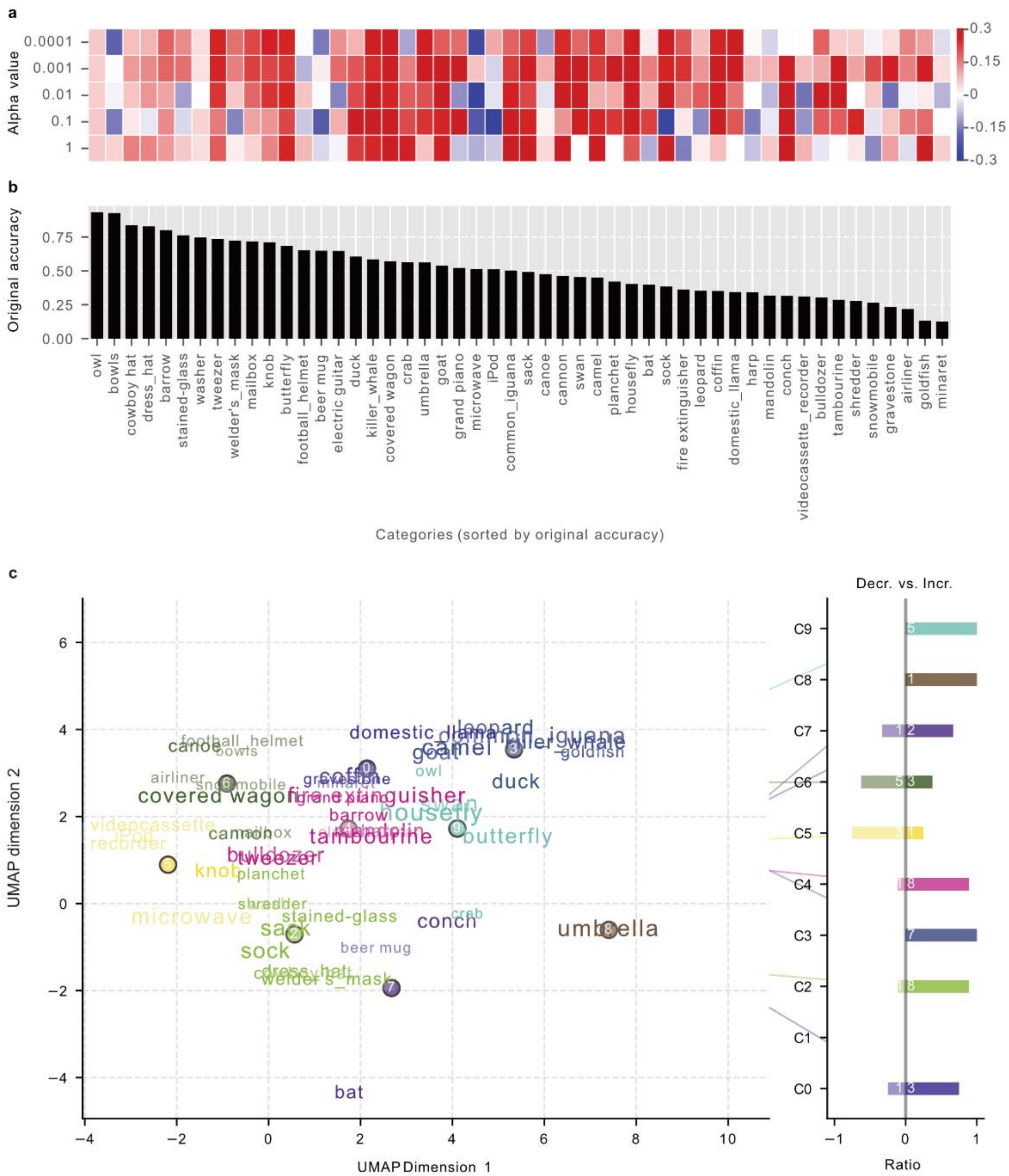

**Supplementary Fig. 11 | Category-specific enhancement patterns of the brain-aligned GloVe vectors in MEG identification. a.** Heatmap displaying the relative changes in the identification accuracy across the brain-alignment parameters (α values) for each test category from ImageNet. The categories are arranged in descending order of the original GloVe identification performance (matching panel b order). The color gradient indicates the performance change relative to the baseline GloVe embeddings. **b.** Baseline identification

accuracy using the original GloVe word embedding vectors for 50 test categories from ImageNet, ranked from highest to lowest performance. The accuracy reflects the zero-shot identification capability on the test dataset. **c.** UMAP projection showing differential category improvements after brain-alignment training. The spatial positioning reflects the whole ImageNet semantic clustering structure (Supplementary Figure 4). The text size represents the magnitude of the accuracy improvement or degradation for the optimal α (which was determined from the fMRI V4 region) compared with the original GloVe vectors. The transparent labels indicate performance decreases, whereas the bold labels indicate accuracy gains. The brain-aligned GloVe vectors demonstrate preferential enhancement for animal categories over inanimate objects in the MEG data, mirroring the pattern observed in the fMRI results.

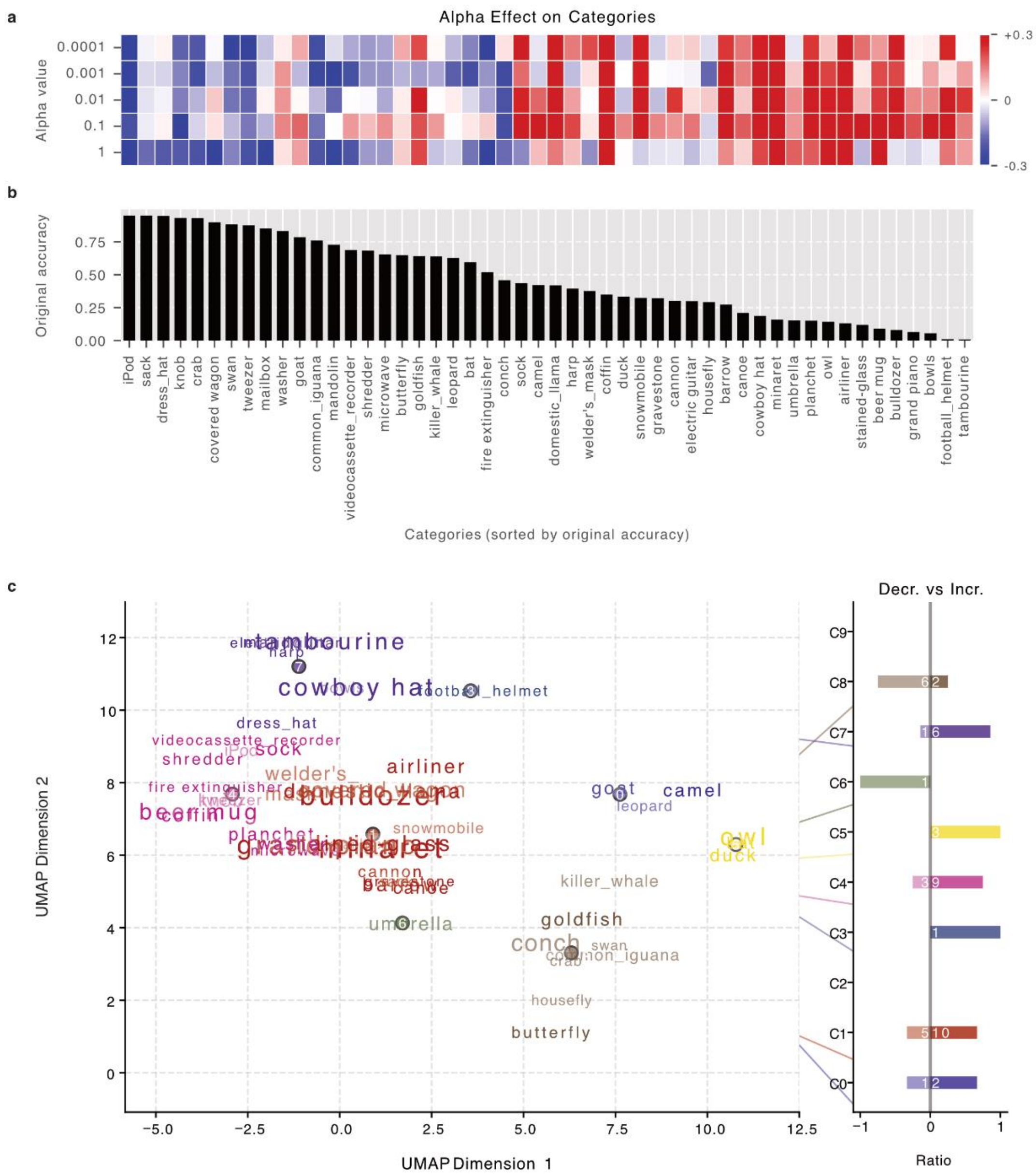

**Supplementary Fig. 12 | Category-specific enhancement patterns of the brain-aligned CLIP vectors in ECoG identification. a.** Heatmap showing the relative change in the identification accuracy across different brain-alignment strengths (α parameters) for each test category from ImageNet. The categories are ordered identically to those in panel b (high to low original accuracy). The color scale represents the accuracy change relative to the original CLIP vectors, with blue indicating decreased performance and red indicating improved performance. **b.**

Baseline identification accuracy using the original CLIP vectors across 50 test categories from ImageNet, sorted from highest (left) to lowest (right) performing categories. The accuracy represents the zero-shot identification performance on the held-out test data. **c.** UMAP visualization of the category-specific improvements following brain-alignment optimization. The categories are positioned according to their location in the whole ImageNet clustering analysis (Supplementary Figure 3). The word size corresponds to the magnitude of the accuracy change for the optimal α value (determined from the fMRI V4 brain region) compared to original CLIP vectors. The transparent text indicates decreased accuracy, whereas the bold text indicates improved accuracy. The brain-aligned CLIP vectors show preferential improvements for object categories compared with biological entities in ECoG data, demonstrating cross-modal consistency with fMRI and MEG findings.

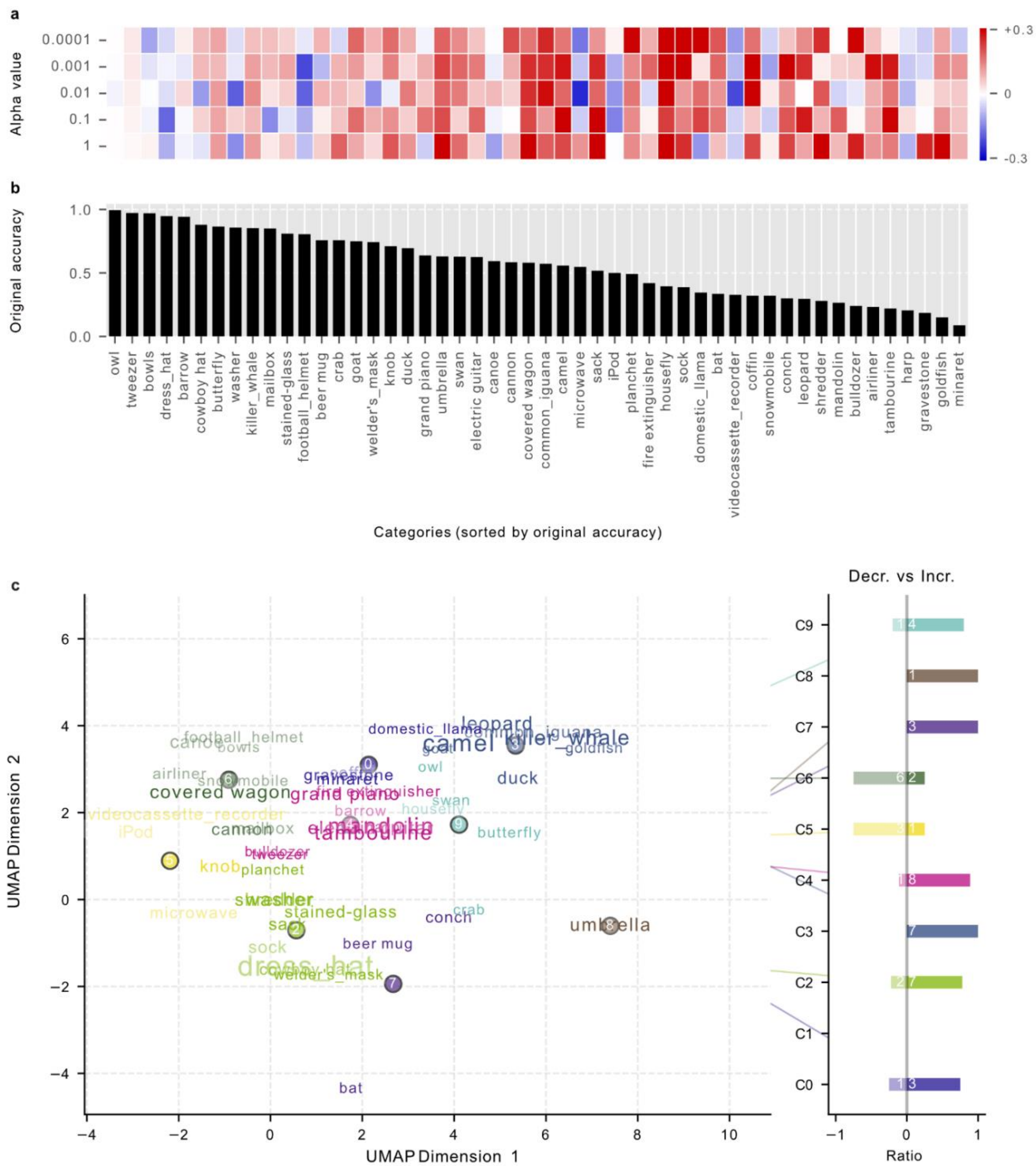

**Supplementary Fig. 13 | Category-specific enhancement patterns of the brain-aligned GloVe vectors in ECoG identification. a.** Heatmap displaying the relative changes in the identification accuracy across brain-alignment parameters (α values) for each test category from ImageNet. The categories are arranged in descending order of the original GloVe identification performance (matching panel b order). The color gradient indicates the performance change

relative to the baseline GloVe embeddings. **b.** Baseline identification accuracy using the original GloVe word embedding vectors for 50 test categories from ImageNet, ranked from highest to lowest performance. The accuracy reflects the zero-shot identification capability on the independent ECoG test data. **c.** UMAP projection showing the differential category improvements after brain-alignment training. The spatial positioning reflects the whole ImageNet semantic clustering structure (Supplementary Figure 4). The text size represents the magnitude of the accuracy improvement or degradation for the optimal α (determined from the fMRI V4 brain region) compared to original GloVe vectors. The transparent labels indicate performance decreases, whereas the bold labels indicate accuracy gains. The brain-aligned GloVe vectors demonstrate preferential enhancement for animal categories over inanimate objects in ECoG data, showing consistent category-selective benefits across neuroimaging modalities.

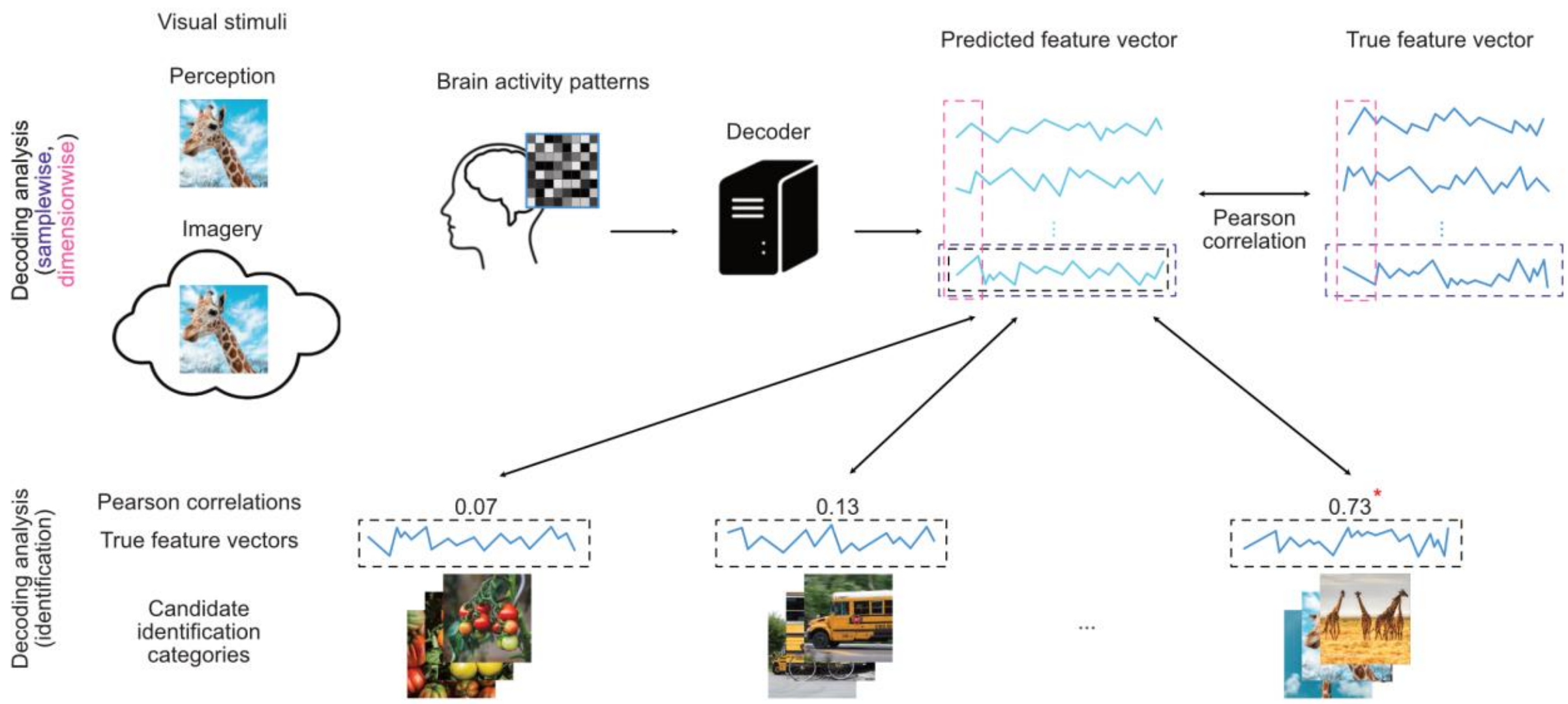

**Supplementary Fig. 14 | Comprehensive evaluation metrics for brain decoding performance.** Brain decoders map neural activity to predicted semantic vectors, which are evaluated using three complementary metrics. **Identification analysis** compares the predicted vectors against all the candidate stimuli using Pearson correlations to determine classification accuracy. **Samplewise analysis** measures the correlation between the predicted and true semantic vectors for each individual stimulus sample, assessing the overall vector reconstruction quality. **Dimensionwise analysis** evaluates the correlation between the predicted and true values for each feature dimension across all samples, measuring the feature-level prediction accuracy. Together, these metrics provide a comprehensive assessment of decoding performance at the categorical, sample, and feature levels.

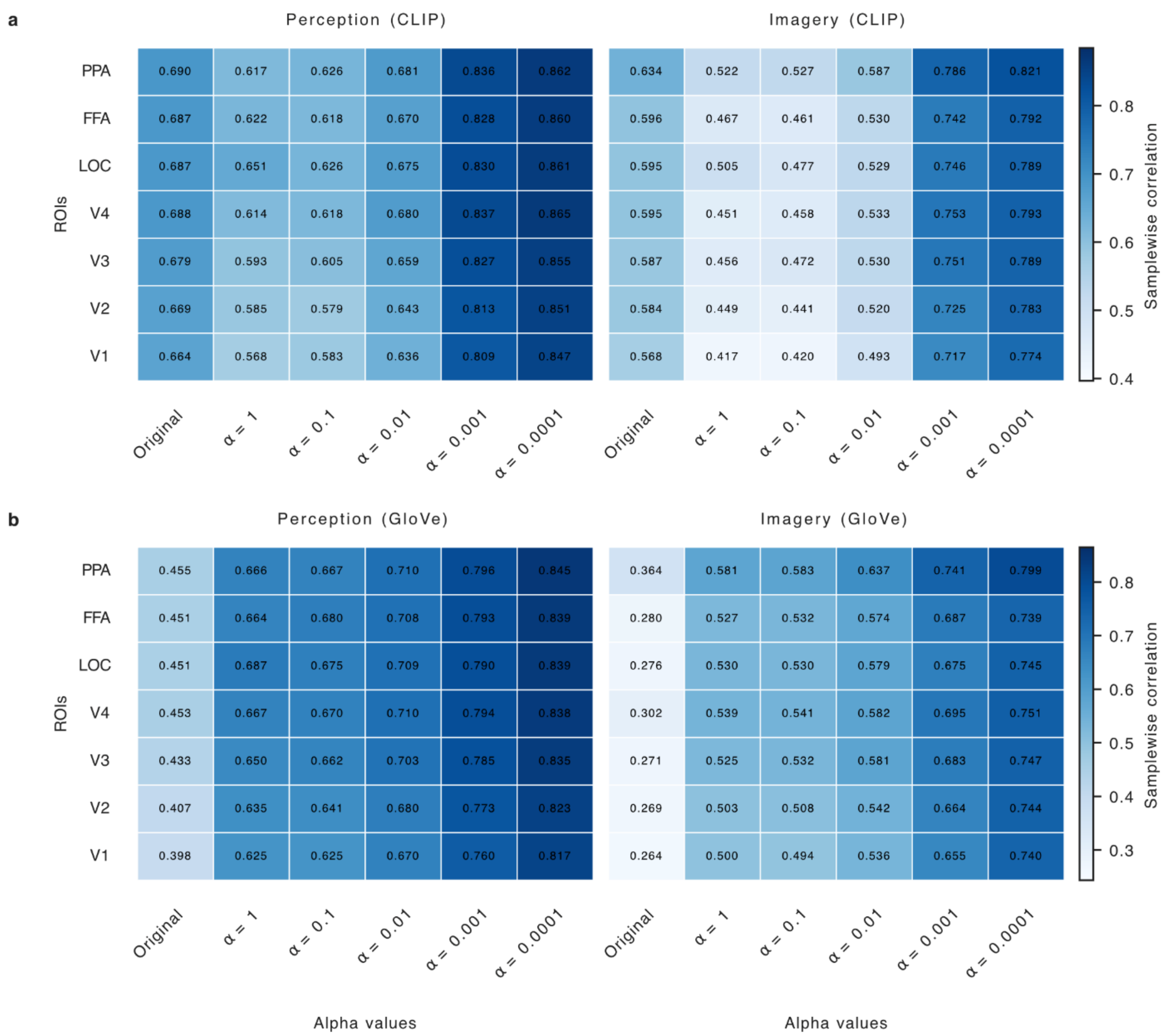

**Supplementary Fig. 15 | Brain-aligned semantic vectors improve the samplewise decoding accuracy across visual cortical regions.** Heatmaps show the mean samplewise correlation between the predicted and true feature vectors for the **a** CLIP-based and **b** GloVe-based semantic representations across seven visual cortical regions (V1, V2, V3, V4, LOC, FFA, and PPA). The left panels show the perception results, and the right panels show the imagery results. The columns represent different levels of brain alignment: "Original" refers to the pretrained feature vectors, whereas the α values (1, 0.1, 0.01, 0.001, and 0.0001) indicate decreasing brain alignment strength. For all the conditions, the real decoding accuracies were significantly greater

than the corresponding shuffled accuracies (one-sided *t* test after Fisher's z-transform, $p < 0.05$; see Supplementary Tables 17–20 for the exact values).

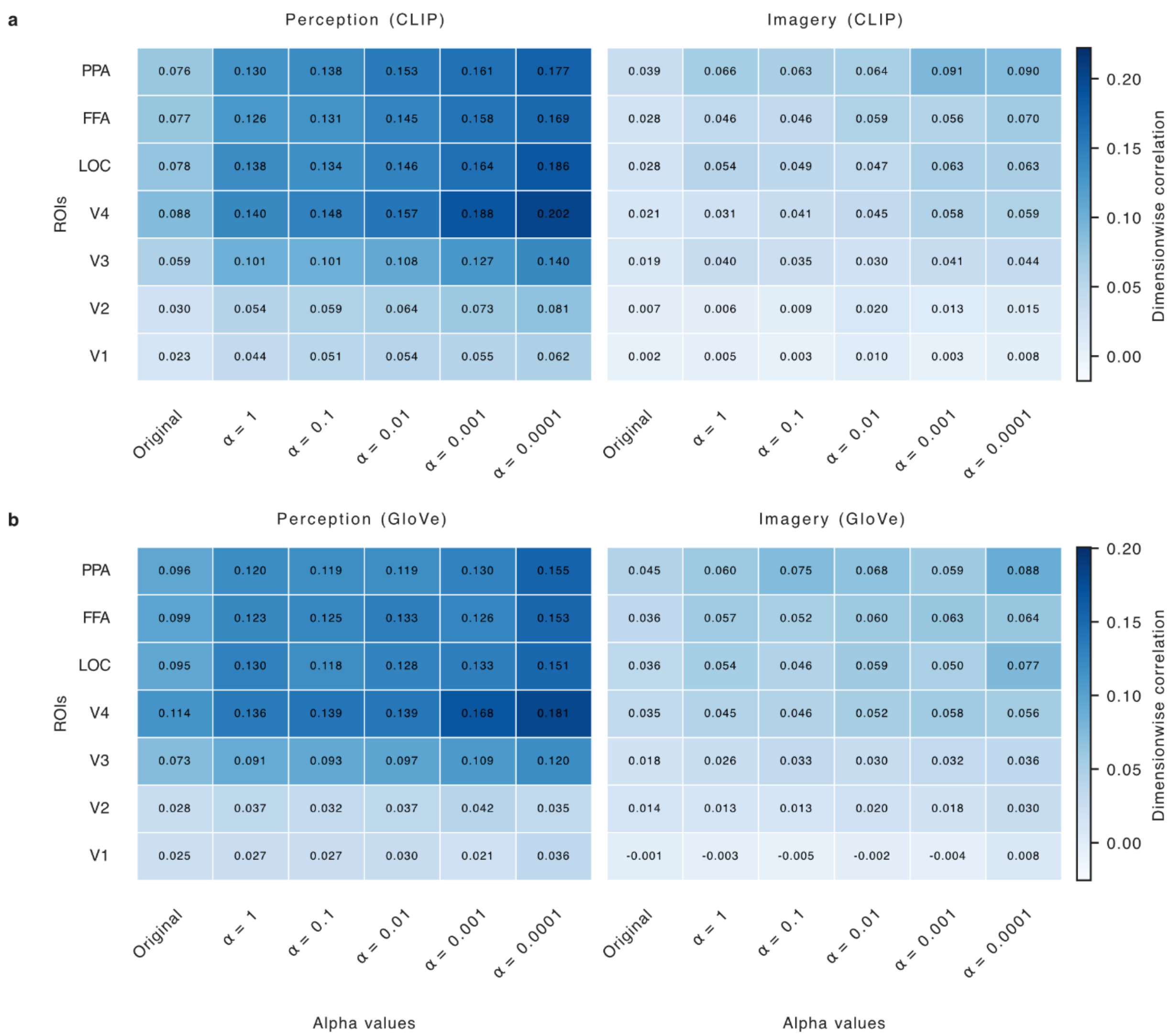

**Supplementary Fig. 16 | Brain-aligned semantic vectors enhance the dimensionwise decoding accuracy with region-specific patterns.** Heatmaps show the mean dimensionwise correlation between the predicted and true feature vectors for the **a** CLIP-based and **b** GloVe-based semantic representations across seven visual cortical regions (V1, V2, V3, V4, LOC, FFA, and PPA). The left panels show the perception results, and the right panels show the imagery results. The columns represent different levels of brain alignment: “Original” refers to the pretrained feature vectors, whereas the α values (1, 0.1, 0.01, 0.001, and 0.0001) indicate decreasing brain alignment strength. The real decoding accuracies were significantly greater than

the corresponding shuffled accuracies for the 42/42 of the CLIP perception data, 30/42 of the CLIP imagery data, 38/42 of the GloVe perception data and 31/42 of the GloVe imagery data (one-sided t test after Fisher's z-transform, $p < 0.05$; see Supplementary Tables 21–24 for the exact values). The higher-order visual areas (V4, LOC, FFA, and PPA) consistently showed greater dimensionwise correlations than the early visual areas did (V1, V2, and V3), with the brain-aligned vectors (particularly $\alpha = 0.001$ and $\alpha = 0.0001$) demonstrating enhanced feature-level decoding performance.

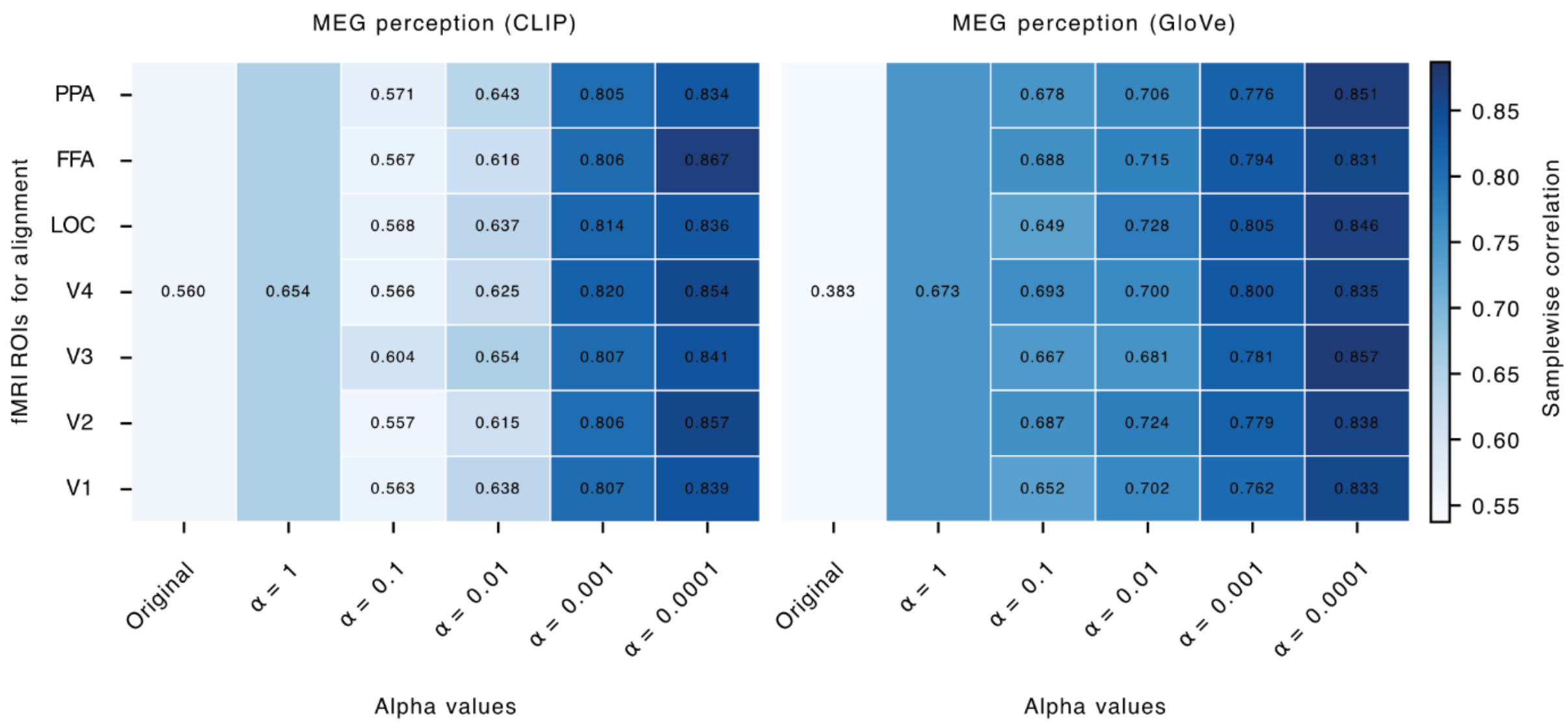

**Supplementary Fig. 17 | Brain-aligned semantic vectors improve the samplewise decoding accuracy in the MEG data.** Heatmaps show the mean samplewise correlation between the predicted and true feature vectors for the CLIP-based (left) and GloVe-based semantic representations (right). The semantic vectors were aligned to seven fMRI visual cortical regions (V1, V2, V3, V4, LOC, FFA, and PPA) and tested for MEG decoding. The columns represent the different levels of brain alignment: "Original" refers to the pretrained feature vectors, whereas the α values (1, 0.1, 0.01, 0.001, and 0.0001) indicate increasing brain alignment strength (lower α = stronger alignment). For all the conditions, the real decoding accuracies were significantly greater than the corresponding shuffled accuracies (one-sided *t* test after Fisher's z-transform, $p < 0.05$; see Supplementary Tables 25–26 for the exact values). Both the CLIP and GloVe representations showed improved MEG decoding performance, with stronger brain alignment to the fMRI ROIs and peak correlations achieved at the lowest α values across all alignment conditions.

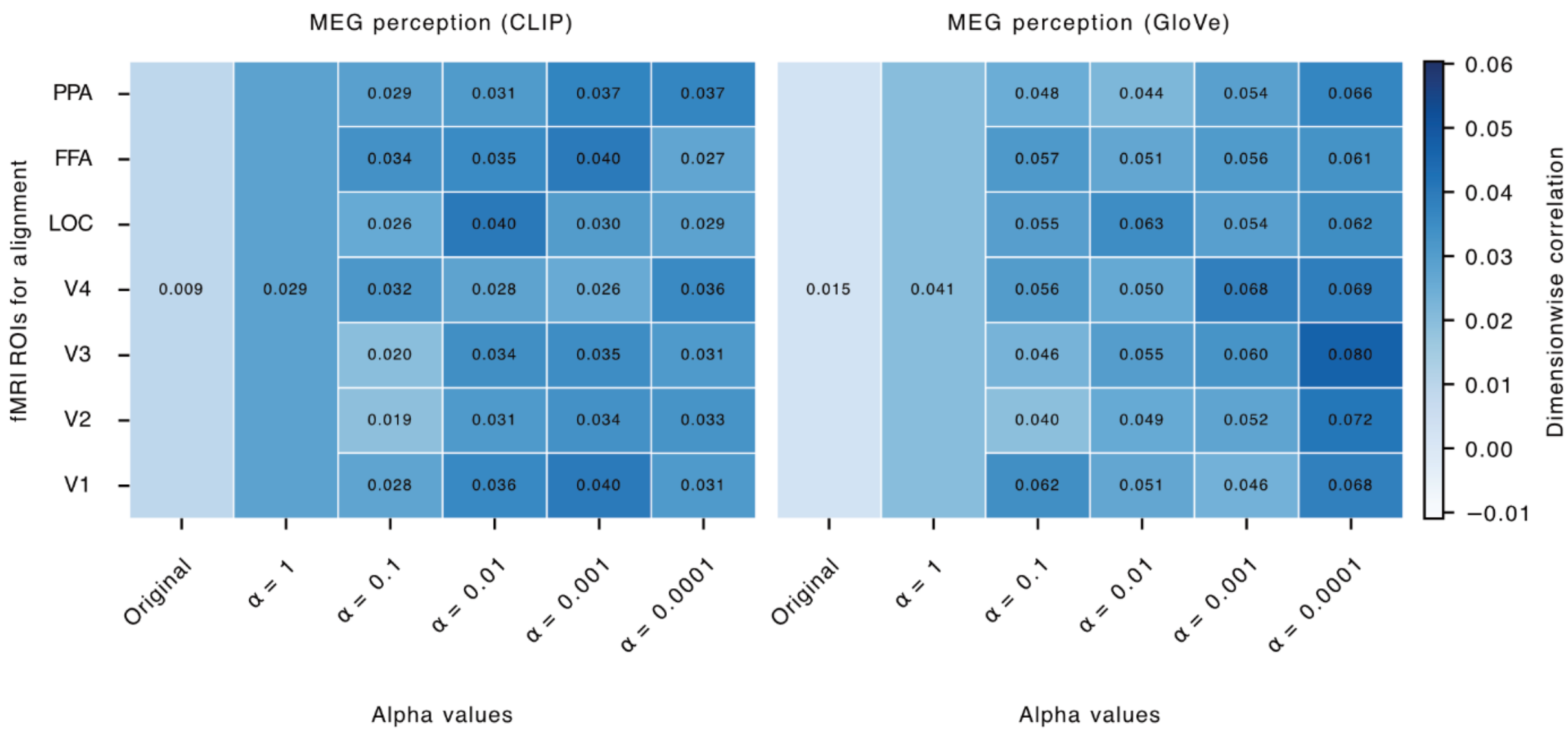

**Supplementary Fig. 18 | Brain-aligned semantic vectors improve the dimensionwise decoding accuracy in MEG decoding.** Heatmaps show the mean dimensionwise correlation between the predicted and true feature vectors for the CLIP-based (left) and GloVe-based semantic representations (right). The semantic vectors were aligned to seven fMRI visual cortical regions (V1, V2, V3, V4, LOC, FFA, and PPA) and tested on the MEG neural data. The columns represent the different levels of brain alignment: “Original” refers to the pretrained feature vectors, whereas the α values (1, 0.1, 0.01, 0.001, and 0.0001) indicate increasing brain alignment strength (lower α = stronger alignment). Statistical significance was assessed using one-sided *t* tests, with significant correlations ($p < 0.05$) observed across multiple alignment conditions and ROIs for both semantic representations (see Supplementary Tables 27–28 for the exact values). The brain-aligned vectors showed enhanced dimensionwise correlations compared to original pretrained vectors, with the GloVe representations demonstrating consistently higher dimensionwise correlations than the CLIP representations across alignment conditions.

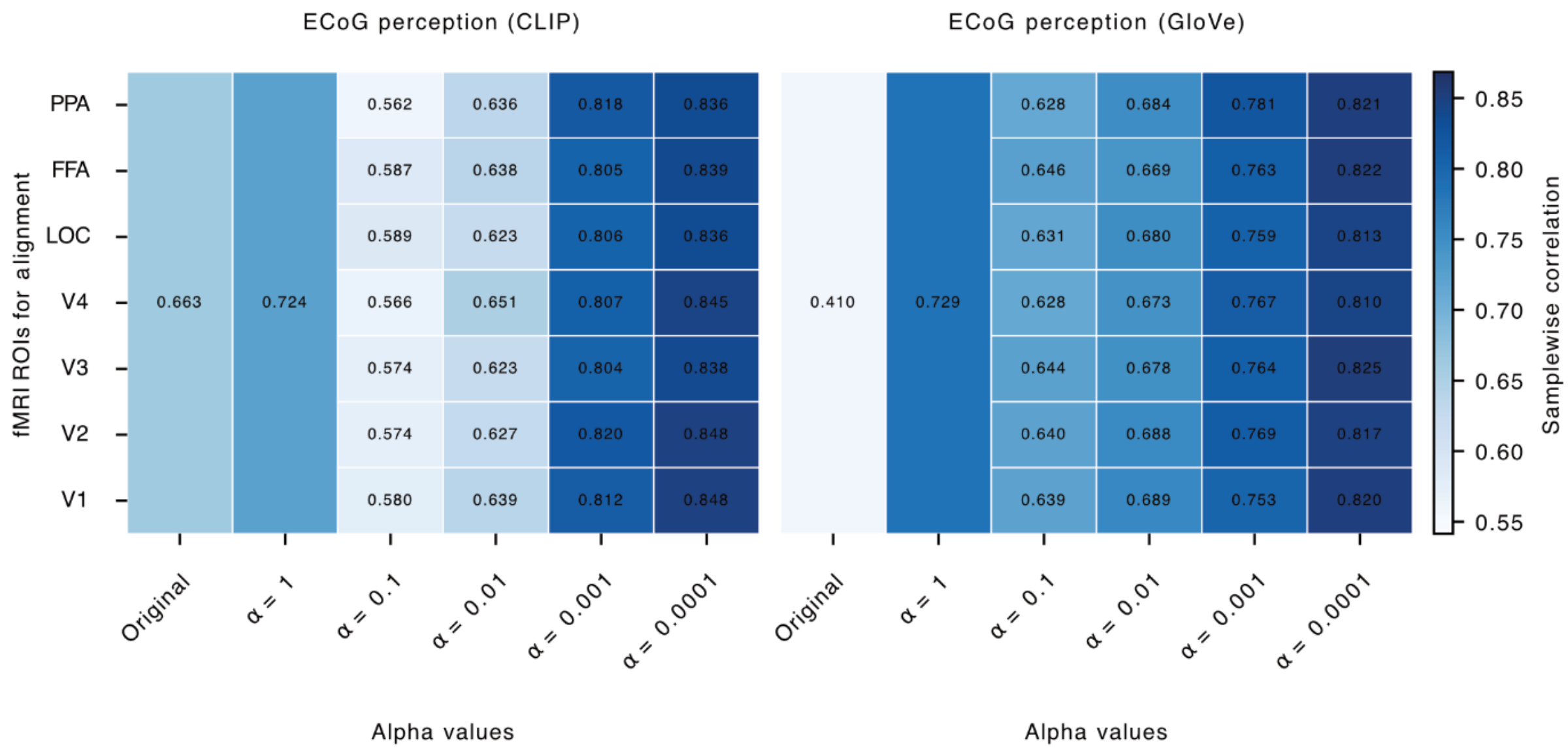

**Supplementary Fig. 19 | Brain-aligned semantic vectors improve the samplewise decoding accuracy in ECoG perception.** Heatmaps show the mean samplewise correlation between the predicted and true feature vectors for the CLIP-based (left) and GloVe-based semantic representations (right) in brain-aligned vectors obtained from different fMRI visual cortical regions (V1, V2, V3, V4, LOC, FFA, and PPA). The columns represent the different levels of brain alignment: "Original" refers to the pretrained feature vectors, whereas the α values (1, 0.1, 0.01, 0.001, and 0.0001) indicate increasing brain alignment strength (lower α = stronger alignment). For all the conditions, the real decoding accuracies were significantly greater than the corresponding shuffled accuracies (one-sided *t* test, $p < 0.05$; see Supplementary Tables 29–30 for the exact values). Both the CLIP and GloVe representations demonstrated substantial improvements in ECoG decoding performance, with stronger brain alignment and peak correlations achieved at the lowest α values across all visual cortical regions.

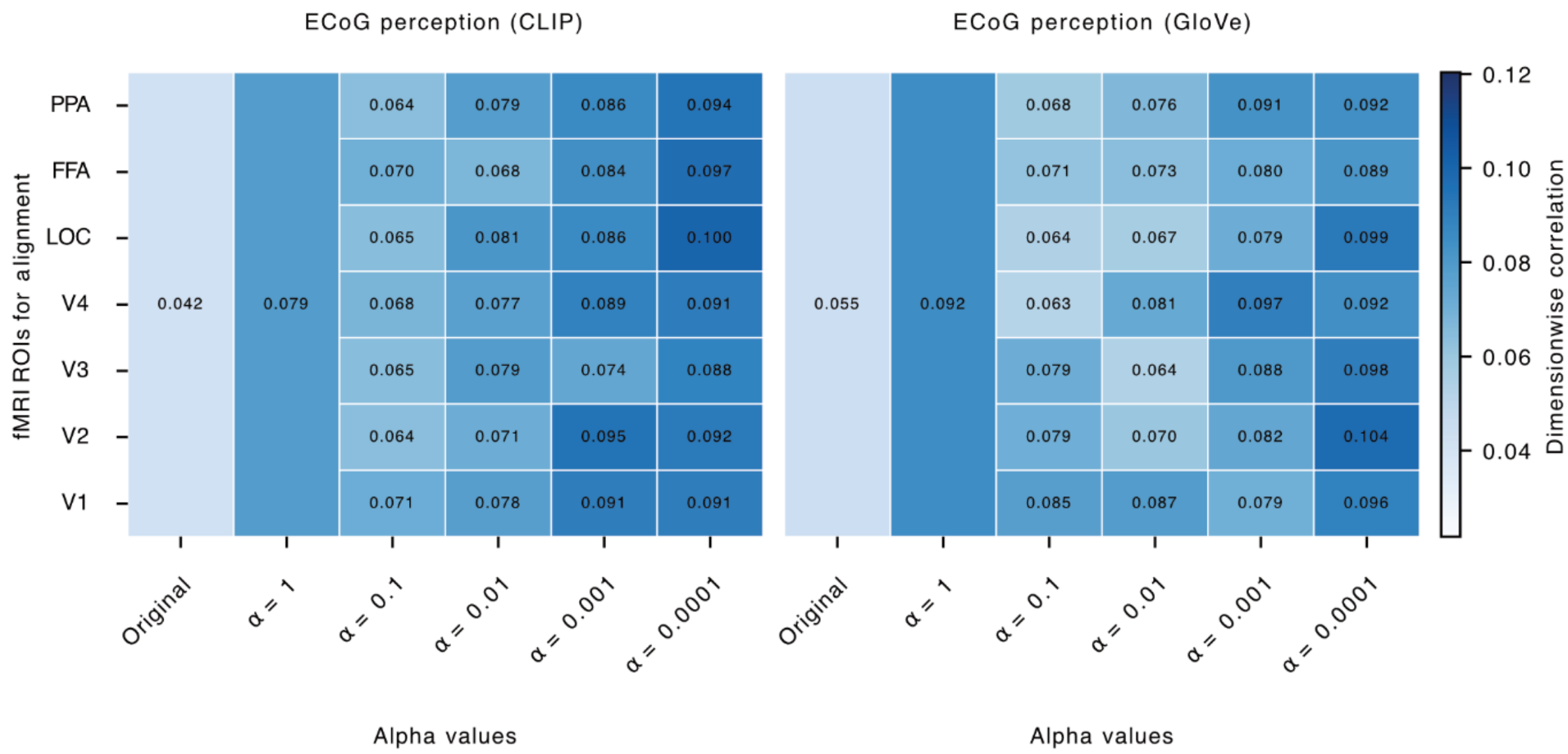

**Supplementary Fig. 20 | Brain-aligned semantic vectors improve the dimensionwise decoding accuracy in ECoG decoding.** Heatmaps show the mean dimensionwise correlation between the predicted and true feature vectors for the CLIP-based (left) and GloVe-based semantic representations (right). The semantic vectors were aligned to seven fMRI visual cortical regions (V1, V2, V3, V4, LOC, FFA, and PPA) and tested on the ECoG neural data. The columns represent the different levels of brain alignment: "Original" refers to the pretrained feature vectors, whereas the α values (1, 0.1, 0.01, 0.001, and 0.0001) indicate increasing brain alignment strength (lower α = stronger alignment). Statistical significance was assessed using one-sided $t$ tests, with significant correlations ($p < 0.05$) observed across multiple alignment conditions and ROIs for both semantic representations (see Supplementary Tables 31–32 for the exact values). The brain-aligned vectors showed enhanced dimensionwise correlations compared to original pretrained vectors.

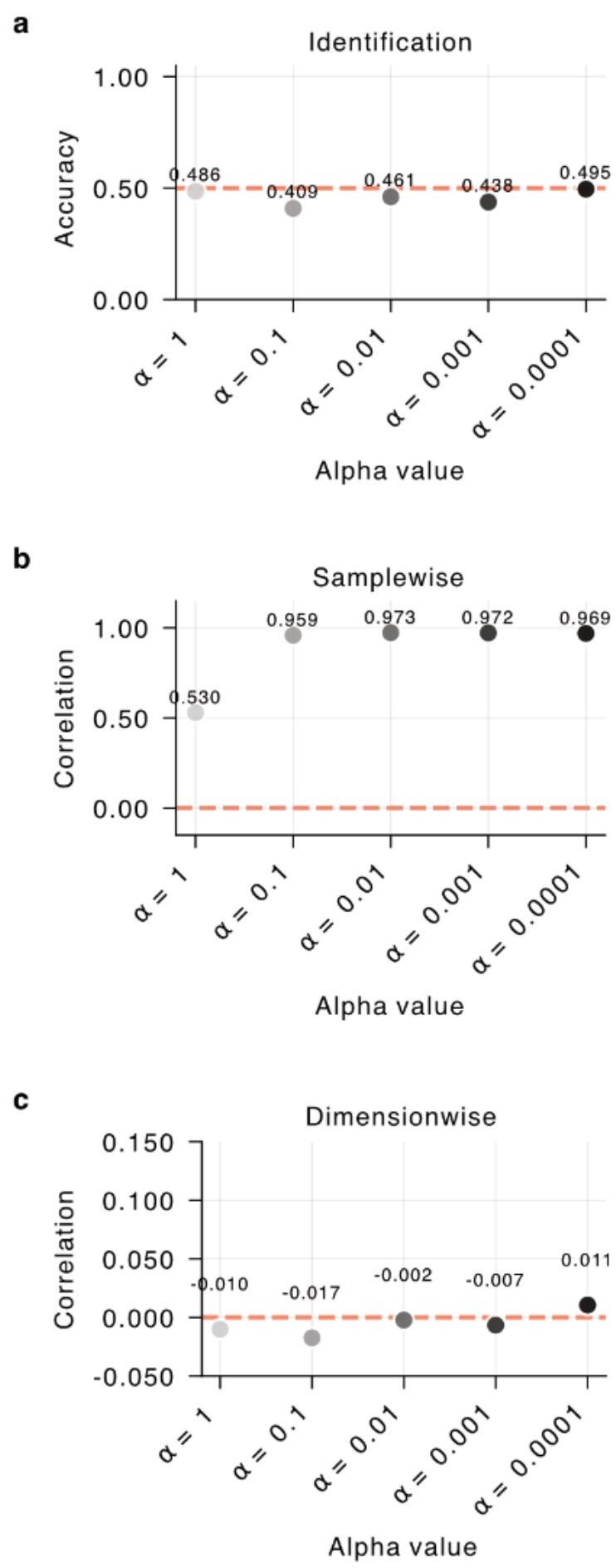

**Supplementary Fig. 21 | Performance of the brain-aligned one-hot vectors.** Decoding performance metrics for the brain-aligned one-hot encoded category representations across different α values for a single subject. **a** Identification accuracy among the brain-aligned one-hot vectors, where all original one-hot labels were passed through the trained model to obtain the brain-aligned representations for the identification candidate set. **b** Samplewise correlation between the predicted and true one-hot vectors. **c** Dimensionwise correlation across the feature dimensions. The original one-hot vectors were excluded from the decoding analysis because their discrete binary structure is incompatible with continuous regression predictions.

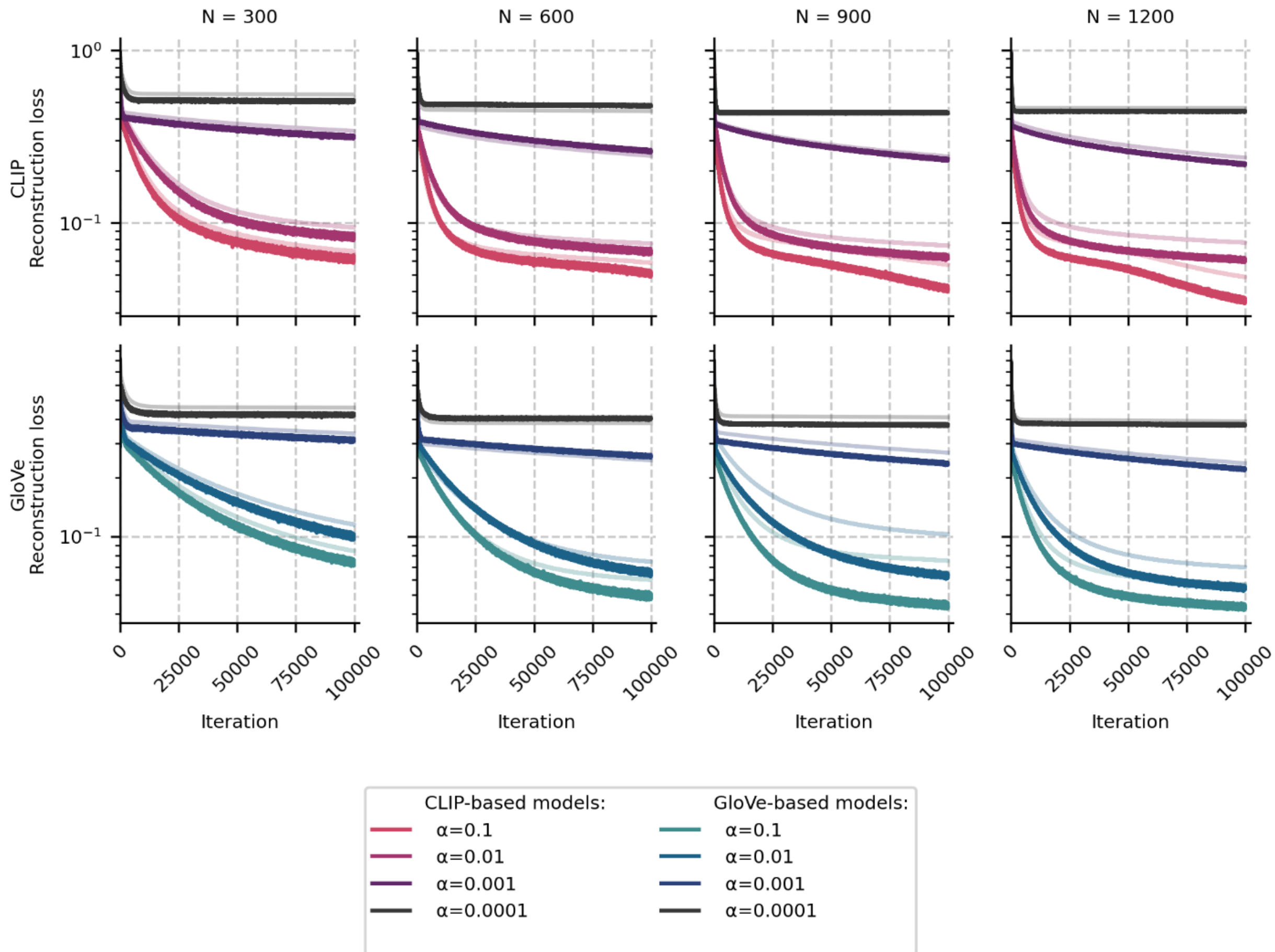

**Supplementary Fig. 22 | Training dynamics for the CLIP-based (top row) and GloVe-based (bottom row) semantic vector reconstruction across varying numbers of brain samples used to train the autoencoder network (N = 300, 600, 900, and 1200 samples)**. The solid lines represent training loss, whereas the transparent lines represent validation loss (15% of the data). The different colors represent different degrees of brain alignment strength. Reconstruction loss (mean squared error) decreases with training iterations, with faster convergence observed when training with larger numbers of brain samples.